\documentclass[aps,pre,reprint,showpacs,superscriptaddress,groupaddress,showkeys,floatfix,byrevtex3]{revtex4-1}
\usepackage{graphicx}
\usepackage{dcolumn}
\usepackage{bm}
\usepackage{srctex}
\usepackage{stackrel}
\usepackage{amsmath}
\usepackage{amsfonts}
\usepackage{amssymb}
\usepackage[dvipsnames]{xcolor}
\usepackage{subfigure}
\usepackage{cancel}
\usepackage{soul}

\graphicspath{{Images/}}

\begin{document}

\title{A Geometrical Method for the Smoluchowski Equation on the Sphere}

\author{Adriano \surname{Vald\'es G\'omez}}
\email[]{adriano@ciencias.unam.mx}
\thanks{author to whom correspondence should be addressed.}

\author{Francisco J. Sevilla}
\email[]{fjsevilla@fisica.unam.mx}
\affiliation{Instituto de F\'isica, Universidad Nacional Aut\'onoma de M\'exico,
Apdo.\ Postal 20-364, 01000, Ciudad de M\'exico, M\'exico}
\date{Today}

\begin{abstract}
A study of the diffusion of a passive Brownian particle on the surface of a sphere and subject to the effects of an external potential, coupled linearly to the probability density of the particle's position, is presented through a numerical algorithm devised to simulate the trajectories of an ensemble of Brownian particles. The algorithm is based on elementary geometry and practically only algebraic operations are used, which makes the algorithm efficient and simple, and converges, in the \textit{weak sense}, to the solutions of the Smoluchowski equation on the sphere. Our findings show that the global effects of curvature are taken into account in both the time dependent and stationary processes. 
\end{abstract}


\maketitle


\section{Introduction}

The diffusion of a tracer particle on the surface of a two-dimensional sphere has served as a simple model to describe many processes observed in nature. On the one hand, such a tracer might correspond to the tip of a vector performing random rotations, where the vector may represent the spin of a molecule, the axis of a rigid body or the direction of motion of a swimming bacteria. On the other, it describes the random motion of particles on the surface of a sphere. In this context, relevant to biophysics, it approximately describes the motion of certain proteins or phospholipids (PIP2 \cite{Fujiwara.2002,Metzler.2016}), on the cell membrane, which are crucial messengers in cell signaling \cite{Haastert.2004, Ma.2004, Meer.2008, Meyers.2006, Hancock.2010}. 

While the mathematical framework that approaches the isotropic rotations of a rigid-body orientation has been developed \cite{FurryPR1957,FavroPR1960,IvanovSovPhysJETP1964,HubbardPRA1972,ValievSovPhysUsp1973,McClungJChemPhys1980}, the Brownian motion of a particle on the surface of a sphere has been analyzed as an extension to general manifolds \cite{GrahamZPhysB1977,vanKampenJStatPhys1986,Risken.1989} of the theory of translational Brownian motion in Euclidean spaces and, at the turn of the 21st century, the improvement on single-particle tracking methods triggered a resurgence of the analysis the random motion of biological tracers on curved surfaces \cite{FaraudoJChemPhys2002}.

Advances on this field have been made for the case when the underlying surface is (nearly) spherical in shape (for instance, the pseudopod formation in cell membranes can be inhibited with a specific hormone \cite{Janetopoulos.2004}). In this case, two mathematical frameworks have been considered to take into consideration the holonomic constraint of moving on the sphere's surface, namely, the Fokker-Planck equation and the Langevin equation, which have also become the theoretical frameworks for studying systems far from thermal equilibrium such as active matter \cite{Castro.2018}, or reaction-diffusion systems \cite{Faustino.2019}.

The free and isotropic motion (without obstacles or external forces) of a particle diffusing on the surface of the two dimensional sphere is well-known, however, the necessity to devise numerical algorithms to generate trajectories on the sphere to investigate the transport processes associated has recently been pointed out \cite{Castro.2014,YangJChemPhys2019,NovikovAppMathComp2020}. A much less investigated case corresponds to that one when the motion is under the influence of external forces. This would be the situation, for instance, when the orientational dynamics of the swimming direction of a bacteria are subject to chemotaxis \cite{SchnitzerPRE1993} or when the rotational Brownian motion is anisotropic \cite{FordPRA1979}.

The general framework to describe Brownian motion constrained to the two dimensional sphere of radius $r$, denoted by $\mathbb{S}^2_{r}$, that is, under a holonomic constraint, is based on the Fokker-Planck equation and has been presented in Refs. \cite{N.G.Van.Kampen.1986,Risken.1989}. In few words, one starts from a formulation of the Fokker-Planck equation in Cartesian coordinates $x^{k}$ ($k=1,2,3$). This is transformed by choosing an appropriate coordinates system (see Appendix \ref{Trans-Fokk}), that allows for the implementation of the constraint. In Cartesian coordinates, the probability density $P(x^{k},t)$ of finding a particle at the coordinates $x^{k}$ at time $t$ satisfies the Fokker-Planck equation
\begin{align}\label{Fokker-Euc}
   [P],_{t}= -  [A^{k} P],_{k} + \frac{1}{2} [B^{k l} P],_{k l},
\end{align}
where $A^{k}$ plays the role of the $k$-th component of a velocity field generated by a force field in the overdamped limit and $B^{kl}= D\, \delta^{k l} $ denotes the diffusion tensor with $D$ the diffusion constant. $[\cdot],_{k}$ and $[\cdot],_{kl}$ denote $ \frac{\partial }{\partial x^k} [\cdot]$ and $ \frac{\partial^2 }{\partial x^{k} x^{l}} [\cdot]$, respectively, and Einstein's summation convention must be understood on repeated indexes. The particle motion is constrained to $\mathbb{S}^2_{r}$ of radius $r$ by requiring that at each instant $x^{k}x_{k}=r^{2}$. Such a constraint is naturally implemented by the use of spherical coordinates: $(x,y,z) \to (r \sin{\theta} \cos{\phi}, r \sin{\theta} \sin{\phi}, r \cos{\theta})$, which induces the transformation $P(x^{k},t)dx^{1}dx^{2}dx^{3}=r^{2}\sin\theta P(\theta,\phi,t)drd\theta d\phi$ and under the constriction we get  $P(x^{k},t)\delta\bigl(\sqrt{x^{k}x_{k}}-r\bigr) \,dx dy dz=\, r^{2}\sin\theta\, P(\theta,\phi, t )d\theta d\phi$. By defining $\overline{P}(\theta,\phi,t)=r^{2}\sin\theta\, P(\theta,\phi, t )$, this satisfies \begin{multline}\label{SmoluOp}
[\overline{P}]_{,t} = -\left [\left ( A^{\theta} + \frac{D}{r^2} \cot{\theta} \right) \, \overline{P}  \right ],_{\theta} - \left [ A^{\phi}  \, \overline{P} \right ]_{,\phi}\\ + \left [\frac{D}{r^2} \, \overline{P} \right ]_{,\theta\theta} + \left [  \frac{D}{r^2 \sin^2{\theta}}  \, \overline{P} \right ]_{,\phi\phi},
\end{multline}
with $A^{\theta}=r^{-1}[\cos\theta\cos\phi\, A^{x}+\cos\theta\sin\phi\, A^{y}-\sin\theta\, A^{z}]$, $A^{\phi}= (r\sin\theta)^{-1} [-\sin\phi\, A^{x}+\cos\phi A^{y}]$. The `spurious' or `noise-iduced' drift, $(D/r^2) \, \cot{\theta}\,\, \overline{P}$ \cite{RyterJMathPhys1980}, appears not only as a consequence of the coordinates transformation, but as part of the total drift, whose effects are physically observable (notice that is proportional to the diffusion constant). For instance, it modifies the `hopping' rate over the potential barriers depending on the particular latitude on the sphere. Indeed, above the equator, the particles are affected by an extra `push' toward the south pole, proportional to the diffusion coefficient. Similarly, below the equator, the extra `push' is toward the north pole. So, curvature will manifest (indirectly) in these kind of physical measurements; for example, if a chemical reaction depends on a potential energy function, the reaction rate will depend on the curvature of the ambient space in which the reaction is taking place.

Exact analytical solutions for \eqref{SmoluOp} are seldom available, except for the case of free diffusion ($A^{\theta}=A^{\phi}=0$), this is why it is important to count on validated numerical methods either to solve it explicitly or implicitly by simulations of such solutions. This becomes relevant especially in the context of applications.

In this work, we devise a numerical algorithm to generate ensemble trajectories of Brownian particles under the effects of an external field, which properly samples the conditional probability density $\bar{P}(\theta,\phi,t | \theta_0,\phi_{0}, t_0)$, solution of equation \eqref{SmoluOp}. We validate the algorithm by comparing its predictions against analytical results in two different scenarios: one considering the time evolution, the other in the stationary regime. In contrast to the algorithm presented in \cite{CarlssonJPhysA2010}, our algorithm is based on the projection over the spherical surface, \textit{after} a proper rescale of the Brownian step computed on the tangent plane. These two steps, rescaling and projecting, implement in an \emph{exact} manner the constriction that maintains the particle motion on the sphere surface, and describe equivalently the diffusion process that underlies equation \eqref{SmoluOp}. Furthermore, there are two main advantages of this method. Firstly, we  avoid the implementation of rotations so we avoid as many trigonometric evaluations as possible; secondly, there is no need of prior knowledge of Riemannian geometry or stochastic calculus to understand its essence, but only elementary geometry.

The paper is organized as follows: in section \ref{NumericAlgorithm} we derive the numerical algorithm that we designed to tackle the problems just defined mathematically in this section. In section \ref{ResultsSec} we present our numerical results as well as the analytical solutions to free diffusion and the solutions to the stationary distributions when there is an external interaction, putting emphasis in the differences with respect to the solutions of flat spaces. We explicitly derive from the associated Fokker-Planck equations the analytic solution to the stationary distributions in the appendix \ref{Deduc-Stationary}. In section \ref{Discuss} we present our concluding remarks and  we leave to appendix \ref{Trans-Fokk} within the appendices, a brief deduction of the transformation rules for the elements of the Fokker-Planck equation that justifies equation \eqref{SmoluOp}.


\section{\label{NumericAlgorithm}The numerical Method}

We present a numerical algorithm that generates an ensemble of trajectories of non-interacting Brownian particles that diffuse on the surface of the sphere $\mathbb{S}^2_{r}$, in the overdamped limit, and under the effects of an arbitrary
external potential. The algorithm is implemented in three dimensions without making use of the standard, but sometimes singular, Langevin-like evolution equations of the spherical angles $\theta$ and $\phi$.

\begin{figure}
\centering
\subfigure[Tangent plane to $\mathbb{S}^2$]{\includegraphics[trim = 10mm 50mm 10mm 30mm, clip,width=1.6in,height=1.9in]{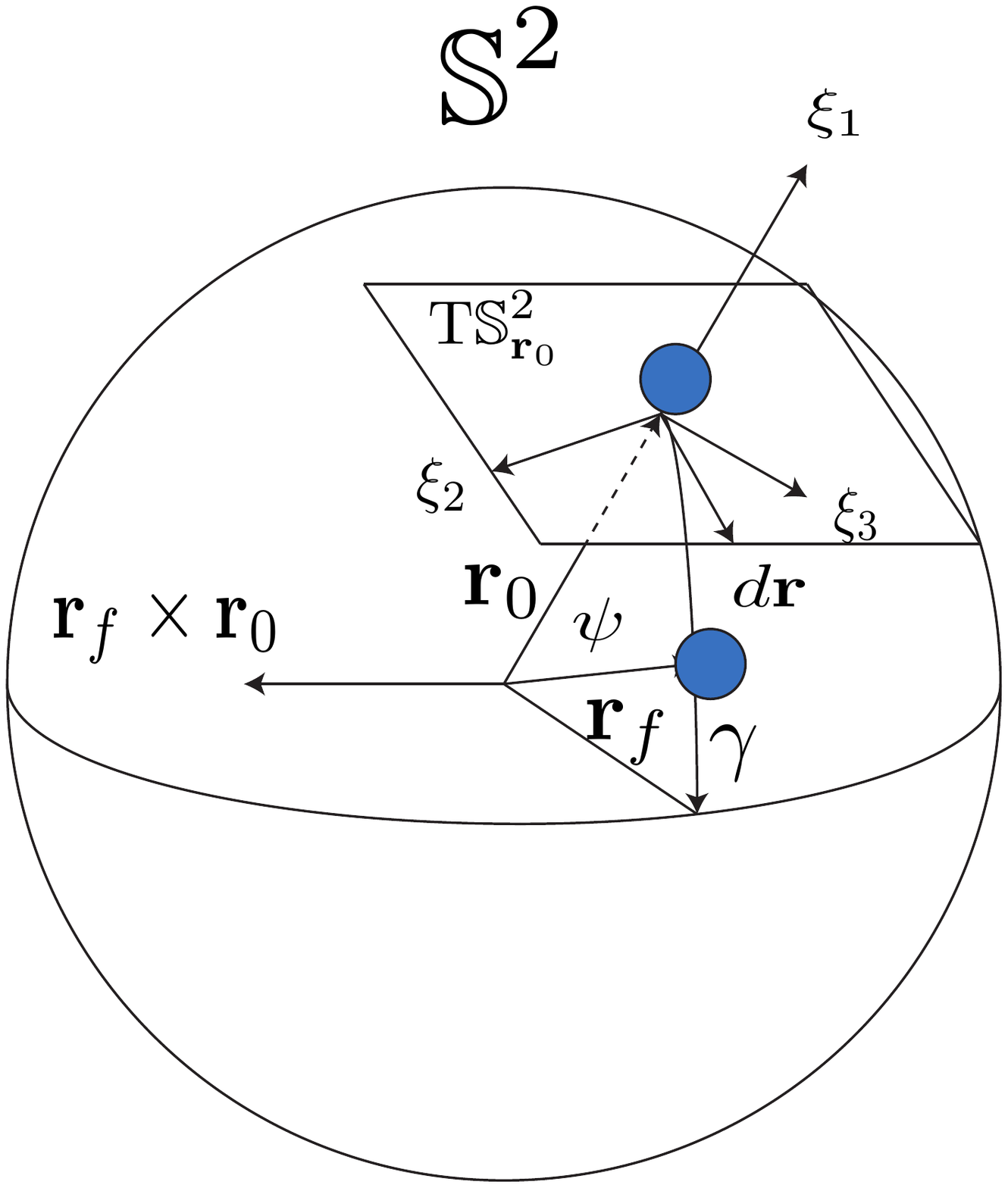}}%
\subfigure[Detail of algorithm step]{\includegraphics[trim = 10mm 30mm 0mm 0mm, clip,width=2.1in,height=2.1in]{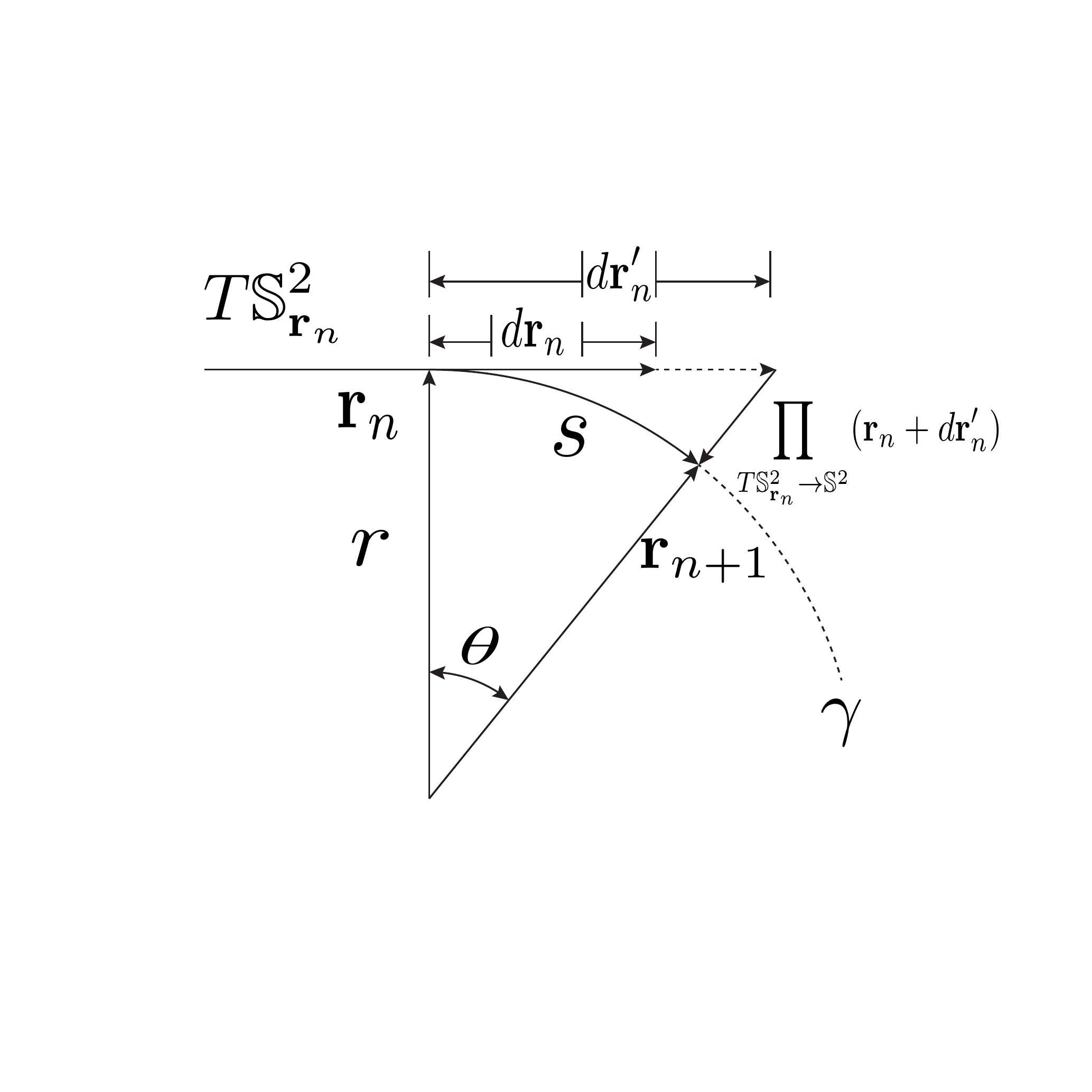}}
\caption{\small $\{\hat{\xi}_1,\hat{\xi}_2,\hat{\xi}_3 \}$ is an orthonormal basis $( \hat{\xi}_{a},\,\hat{\xi}_{b}  ) = \delta_{a,b}$ in $\mathbb{R}^3$, such that 
$\{\hat{\xi}_2,\hat{\xi}_3\}$ is a basis in the tangent space $\mbox{T}_{\mathbf{r}_0}\mathbb{S}^2$ to the sphere $\mathbb{S}^2$, at $\mathbf{r}_0 \in \mathbb{S}^2$. 
In this image $\mathbf{r}_{f}$ is the final vector position after the displacement over a geodesic $\gamma$.}
\label{Algoritmo_01}
\end{figure}

The instantaneous particle's position at time $t$, with respect to the sphere center, is denoted with $\boldsymbol{r}(t)=r\hat{\boldsymbol{n}}(t) $ with $\hat{\boldsymbol{n}}=\bigl(\sin\theta(t)\cos\phi(t),\sin\theta(t)\sin\phi(t),\cos\theta(t)\bigr)$ $\in \mathbb{R}^3$. At this point on the surface,  the tangent plane 
$\text{T}_{\boldsymbol{r}(t)}\mathbb{S}^{2}$  is used to approximate the particle position after a time interval $d t$ namely, $\boldsymbol{r}(t+d t) = \boldsymbol{r}(t) + d\boldsymbol{r}(t)$, i.e. the new approximated particle position is computed locally  
by calculating the net change $d\boldsymbol{r}(t)=d\boldsymbol{r}_{n}(t)+d\boldsymbol{r}_{f}(t)$, where $d\boldsymbol{r}_{n}(t)$ is the change due to the effects of noise, and $d\boldsymbol{r}_{f}(t)$ the change caused by all of the forces on the particle as is shown schematically in figure \ref{Algoritmo_01}. 

The noisy term $d\boldsymbol{r}_{n}(t)$ is computed by generating two pseudo-random numbers: one normally distributed using the \textit{Mersenne Twister method}, and the other uniformly  distributed in $[0,2\pi]$. In the finite time interval $\Delta t$ we have that
\begin{subequations}\label{Tangent-Forces}
\begin{align}\label{Deltarf}
d\boldsymbol{r}_{n}=& \sqrt{4D\Delta t}\, |W| \left( \cos{[\Psi]}\, \mathbf{\hat{\xi}}_2+ \sin{[\Psi]}\, \mathbf{\hat{\xi}}_3 \right ),
\end{align}
where $W$ is a random variate drawn from the normal distribution, $\mathcal{N}(0,1)$, with vanishing mean and variance $1$, $D$ being the diffusion coefficient. $\Psi$ is a uniformly random variate in $[0,2\pi]$ if we take the absolute value of $W$, or in $[0,\pi]$ if we let $W$ take on negative values as well. These are two equivalent ways to generate a two-dimensional normal distribution on the tangent planes to the sphere. $\hat{\xi}_{2}$ and $\hat{\xi}_{3}$ form a orthonormal basis for $\text{T}_{\boldsymbol{r}(t)}\mathbb{S}^{2}$, which due to the statistical nature of $\Psi$, we can chose arbitrarily (this is how we employed it in the case of free diffusion on the sphere, see the code in the Github site \cite{GitHub_Repo}).
$d\boldsymbol{r}_{f}$ on the other hand, is computed simply by projecting the deterministic forces (following D'Alambert's principle) onto $\text{T}_{\boldsymbol{r}(t)}\mathbb{S}^{2}$ and updating using the Euler method, namely 
\begin{equation}\label{Deltadf}
d\boldsymbol{r}_{f}= \zeta^{-1} \Delta t\left[\mathcal{A}^{\theta}\, \hat{\boldsymbol{\theta}}_{\boldsymbol{r}}(t)+\mathcal{A}^{\phi}\, \hat{\boldsymbol{\phi}}_{\boldsymbol{r}}(t)\right],
\end{equation}
\end{subequations}
where $\mathcal{A}^{\theta}$ and $\mathcal{A}^{\phi}$ are the components of the deterministic forces 
along the directions given by the unit vectors $\hat{\boldsymbol{\theta}}_{\boldsymbol{r}}(t)=\bigl(\cos\theta(t)\cos\phi(t),\cos\theta(t)\sin\phi(t),-\sin\theta(t)\bigr)$ and $\hat{\boldsymbol{\phi}}_{\boldsymbol{r}}(t)$ $ =\bigl(-\sin\phi(t),\cos\phi(t),0\bigr)$ at position $\boldsymbol{r}(t)$, respectively. $\zeta$ in equation \eqref{Deltadf} denotes the damping constant that emerges from the coupling between the particle motion and the thermal bath, which gives rise to the fluctuating motion of the particle.
Since $\boldsymbol{r}(t+\Delta t) =\boldsymbol{r}(t)+d\boldsymbol{r}$ does not lie on $\mathbb{S}^{2}_{r}$, but on $\text{T}_{\boldsymbol{r}(t)}\mathbb{S}^{2}$, it must to be projected back to the sphere surface (see figure \ref{Algoritmo_01}). This can be accomplished by the implementation of the following algebraic scheme that allows an accurate calculation of the particle displacement on the sphere. First, once $d\boldsymbol{r}$ is computed, it is scaled by a factor $\alpha$ in such a way that the arc length of the geodesic $\gamma$ between $\boldsymbol{r}(t)$ and the projection of $\boldsymbol{r}(t)+\alpha d\boldsymbol{r}$ on the sphere (see figure \ref{Algoritmo_01}b), denoted with $\boldsymbol{r}_{_{\Pi}}(t+\Delta t)$, coincides with $\vert\vert d\boldsymbol{r}\vert\vert$. From simple geometry it can be shown that  $\alpha=r\tan(\vert\vert d\boldsymbol{r}/r\vert\vert)/\vert\vert d\boldsymbol{r}\vert\vert$. Thus, the particle position is updated according to 
\begin{multline}\label{Algo-Field}
\boldsymbol{r}_{_{\Pi}}(t +  \Delta t) = \prod_{T \mathbb{S}^2_{\mathbf{r}(t)} \to \mathbb{S}^2} \Biggl[ \boldsymbol{r}(t)  + r \tan \biggl \{ \biggl|\biggl| \frac{2\sqrt{D\Delta t}\, \vert W\vert}{r} \times\\
\Bigl(\cos[\Psi]\, \mathbf{\hat{\xi}}_2(t) + \sin{[\Psi]}\, \mathbf{\hat{\xi}}_3(t) \Bigr) \\
+  \frac{ \Delta t}{r \zeta} \biggl( \mathcal{A}^{\theta} \, 
\mathbf{\hat{\boldsymbol{\theta}}}_{\boldsymbol{r}}(t) +  \mathcal{A}^{\phi} \, \mathbf{\hat{\boldsymbol{\phi}}}_{\boldsymbol{r}}(t) \biggr) \biggr|\biggr| \biggr\} \hat{d\boldsymbol{r}} \Biggr], 
\end{multline}
which forms the basis of our numerical algorithm that takes into account the effects of the external forces. In \eqref{Algo-Field}, the symbol $\prod_{\mathbb{S}^2}(\mathbf{x})$ denotes the projection operator $\boldsymbol{x}( \boldsymbol{x}, \boldsymbol{x} )^{-1/2}$, with $(\,,\,)$ the \textit{natural scalar product} in $\mathbb{R}^3$, used to bring back the particles from $\mbox{T}_{\boldsymbol{x}}\mathbb{S}^2\subset \mathbb{R}^3$ onto $\mathbb{S}^2$; and $\hat{d\boldsymbol{r}}=d\boldsymbol{r}/\vert\vert d\boldsymbol{r}\vert\vert$. 
In the absence of noise, equation \eqref{Algo-Field} reduces to the integration of the deterministic part of the dynamics, i.e. 
\begin{align}
\boldsymbol{r}_{_{\Pi}}(t+\Delta t) = \prod_{T \mathbb{S}^2_{\boldsymbol{r}(t)} \to \mathbb{S}^2} \left [ \boldsymbol{r}(t) + r \tan{ \left \{ || d\boldsymbol{r}_{f}||/r \right \} \hat{d\boldsymbol{r}}_f} 
\right ].
\end{align}
The method is of first order in $\Delta t$ in the context of stochastic differential equations, however 
it could be possible to develop an analogous form of the \textit{the Miltein's method} \cite{Higham.2001}, which is based on an It\^{o}-Taylor expansion. Some of these topics are discussed in \cite{Gardiner.1997}.

In the absence of external forces, the vector 
\begin{equation}
\prod_{T \mathbb{S}^2_{\boldsymbol{r}(t)} \to \mathbb{S}^2}\bigl[\boldsymbol{r}(t)+r \tan{ \left \{ || d\boldsymbol{r}_{n}||/r \right \} \hat{d\boldsymbol{r}}_n}\bigr],
\end{equation}
is the one that takes into consideration the particle's Brownian motion on the sphere, where $d\boldsymbol{r}_{n}$ is given in equation \eqref{Deltarf}. As such, and as our numerical analysis shows afterwards, the underlying geometry of this updating rule sheds light over the geometric meaning of the three terms proportional to $D$ in the Smoluchowski equation \eqref{SmoluOp}, since these same terms fully describe the diffusive aspects of Brownian motion on the sphere. We want to emphasize that our algebraic scheme, the combination of scaling and projecting after updating in the tangent plane, is equivalent to the exact updating rotation of $\boldsymbol{r}(t)$ and therefore no additional systematic error is introduced in such procedure (see \cite{SupMat}).

\section{Results}\label{ResultsSec}

In this section, we compare the results obtained from an statistical analysis based on an ensemble of trajectories that start at $\theta(t=0)=0$ (north pole) computed from the algorithm described in the last section. Firstly, we perform a comparison in the context of free diffusion on $\mathbb{S}^{2}_{r}$, for which an analytic solution for the marginal probability distribution $\overline{P}(\theta,t|0,0)=2\pi \overline{P}(\theta,\phi,t|0,0,0)$ is available at all times, namely \cite{Roberts.1960,Hobson.1965, Asmar.2005}
\begin{align}\label{Fokker-Planck-Sol}
\overline{P}(\theta, t |0,0)= \sum_{n=0}^{\infty} \frac{2n+1}{2\pi r^2}P_{n}(\cos{\theta})e^{-n(n+1)D\,t/r^2 }\sin{\theta},
\end{align}
where the spherical addition theorem \cite{Jackson.1999} is used to write the sum of the product of spherical harmonics as a Legendre polynomial $P_{n}(\cos\theta )$. We focus on the standard parameters that characterize the particle diffusion, namely, the mean square displacement and the position autocorrelation function $\langle \hat{\boldsymbol{n}}(t) \cdot  \hat{\boldsymbol{n}}(0) \rangle=\langle\cos\theta(t)\rangle=\langle P_{1}(\theta)\rangle$. Furthermore, we calculate the complete histogram of the particle positions and the mean value of the polar angle, $\langle\theta\rangle$. 

Although the quantities of interest have an analytical expression obtained from the solution \eqref{Fokker-Planck-Sol} of the Fokker-Planck equation \eqref{SmoluOp} in the absence of external fields, only the position autocorrelation function can be written in a closed form, namely
\begin{align}\label{AutoCorrFunct}
 \langle \hat{\boldsymbol{n}}(t) \cdot  \hat{\boldsymbol{n}}(0) \rangle = \exp{\left [-2Dt/r^2  \right ]}.
\end{align}
For the calculation of the mean polar angle $\langle\theta(t)\rangle=\int_{0}^{\theta}d\theta\,\theta \overline{P}(\theta,t\vert0,0)$ and the mean square displacement $\langle\boldsymbol{r}^{2}(t)\rangle$ (which under the initial conditions coincides with $\langle\theta^{2}(t)\rangle=\int_{0}^{\theta}d\theta\,\theta^{2} \overline{P}(\theta,t\vert0,0)$) we used a numerical evaluation of the involved integrals using \eqref{Fokker-Planck-Sol}.

\begin{figure}[ht]
\includegraphics[trim = 70mm 110mm 0mm 100mm, clip, width=\columnwidth, height=25mm]{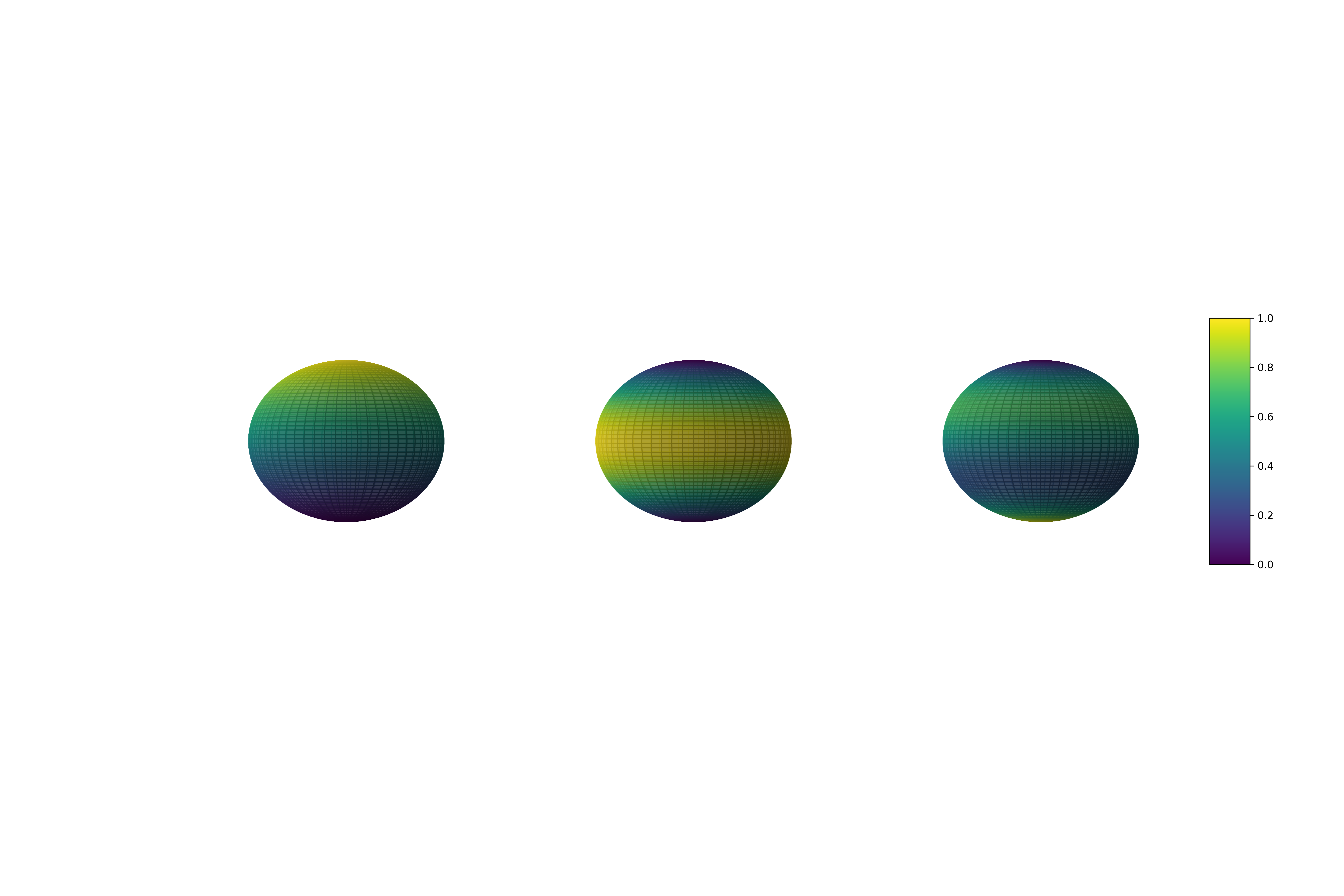}
\caption{Heatmaps for the three different potentials in equation \eqref{PotentialSet}. (Left) $U_{\lambda}(\theta) = \lambda\, Y_{1}^{0}$, (middle) $U_{\lambda}(\theta) = -\lambda\, Y_{2}^{0}(\theta,\phi)$, and (right) $U_{\lambda}(\theta) =- \lambda\, Y_{3}^{0}(\theta,\phi)$. Bright zones correspond to high values of the potentials whereas dark regions correspond to small values.}
\label{3-Potentials}
\end{figure}

Afterwards, we use our numerical algorithm to analyze the of diffusion of particles on $\mathbb{S}^{2}$ in the presence of an external field which in general can be expanded in the spherical harmonics $Y_{l}^{m}(\theta,\phi)$. For the sake of a clear analysis, we consider an external potential $U_{\lambda}$ that depends only on $\cos\theta$ \big(or equivalently that depends only on the quantity $z/(x^{2}+y^{2}+z^{2})^{1/2}$\big), thus focusing on the cases with azimuthal symmetry. $\lambda$ denotes the external potential strength whose physical dimensions are of energy. In such a case we have that the components of the force field in spherical coordinates are given by
\begin{subequations}
\begin{align}
\mathcal{A}^{x}_{\lambda}=&\frac{1}{r}U^{\prime}_{\lambda}(\cos\theta)\cos\theta\sin\theta\cos\phi,\\
\mathcal{A}^{y}_{\lambda}=&\frac{1}{ r}U^{\prime}_{\lambda}(\cos\theta)\cos\theta\sin\theta\sin\phi,
\end{align}
where $U^{\prime}_{\lambda}(w)=\frac{d}{dw}U_{\lambda}(w)$. Thus we get $\mathcal{A}^{\phi}_{\lambda}=0$. Likewise,  
\begin{equation}
\mathcal{A}^{z}_{\lambda}=-\frac{1}{ r}U^{\prime}_{\lambda}(\cos\theta)\sin^{2}\theta,
\end{equation}
and therefore (see Appendix \ref{Trans-Fokk})
\begin{equation}\label{ForceTheta}
\mathcal{A}^{\theta}_{\lambda}=\frac{1}{ r}U^{\prime}_{\lambda}(\cos\theta) \sin\theta.
\end{equation}
\end{subequations}
That being so, $U_{\lambda}(\theta)$ can be expanded in the Legendre polynomials $P_{l}(\cos\theta)$. We chose three specific cases for $U_{\lambda}(\theta)$, namely 
\begin{subequations}\label{PotentialSet}
\begin{align}
U_{\lambda}(\theta) &= \lambda\, Y_{1}^{0}(\theta,\phi)=\lambda\, \sqrt{\frac{3}{4\pi}}\, \cos{\theta} ,\label{U1}\\ 
U_{\lambda}(\theta) &= -\lambda\, Y_{2}^{0}(\theta,\phi)=\lambda\, \sqrt{\frac{5}{16\pi}}\, \left(1-3\cos^2{\theta}\right ),\label{U2}\\ 
U_{\lambda}(\theta) &= -\lambda\, Y_{3}^{0}(\theta,\phi)= \lambda\,  \sqrt{\frac{7}{16\pi}}\,\cos\theta\left( 3 -5\cos^2{\theta}  \right),\label{U3}
\end{align}
\end{subequations}
which are depicted in figure \ref{3-Potentials}. 
The (minus) sign in \eqref{U2} and \eqref{U3} is chosen in order to have a potential with local minima at $\theta=0,\, \pi$ and one local maximum at $\theta=\pi/2$ in the first case. For the second case, we have a local minimum at $\theta=0$ and a maximum at $\theta=\pi$ separated by a local minimum at $\theta=\arccos\left(1/\sqrt{5}\right)$. 

As has been pointed out before, there are no available analytical solutions to the time dependent problem, but there are for the stationary solutions. The solution corresponding to \eqref{U3} cannot be expressed in terms of elementary functions, so we will evaluate it numerically. We use as the characteristic length the radius of the sphere, such that when combined with the diffusion coefficient it defines the time scale $\tau = r^2/D$. Except when is stated otherwise, we have used $D = 0.1 $, $\lambda \approx 2.24$, and $\Delta t = \ln{2} \times 10^{-3} \tau$ for the time step.

\subsection{Free diffusion}\label{FreeDiffSec}

\begin{figure}[ht]
\centering
\includegraphics[width=3.6in]{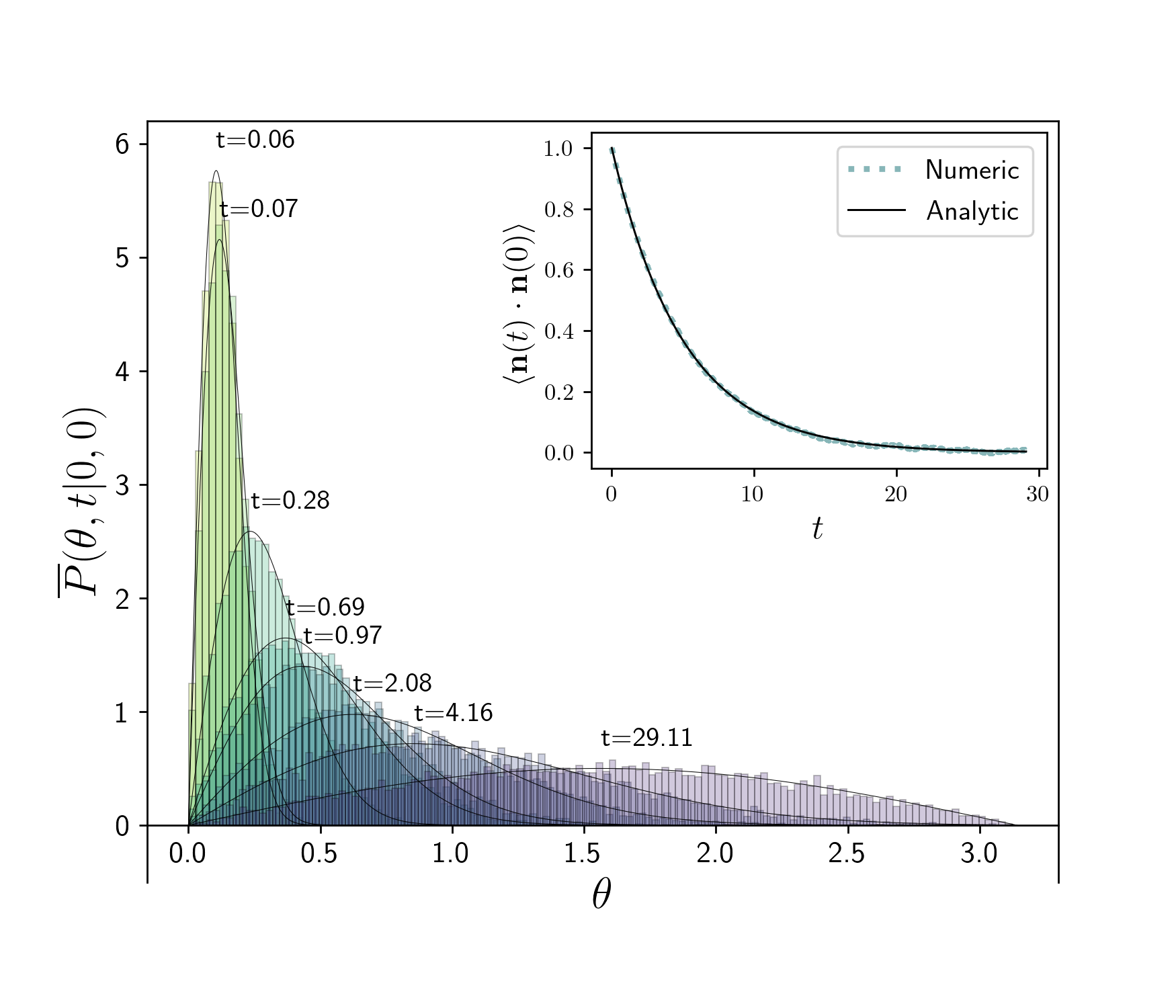}
\caption{Comparison between the (normalized) histogram of the theta variable $\theta$ obtained from the ensemble of 10$^{4}$ trajectories calculated with our numerical algorithm (shaded-color bars), and the probability density $\overline{P}(\theta,t\vert0,0)$ given by Eq. \eqref{Fokker-Planck-Sol} (solid lines), at different times. (Inset) Comparison between the time dependence of the autorcorrelation function \eqref{AutoCorrFunct} (solid line) and the corresponding one computed from the ensemble of trajectories (blue dots).}
\label{Histograma_Theta}
\end{figure}

In this section we show the results obtained from our devised numerical algorithm of the time evolution of ensembles of independent particles that start at the north pole. Firstly, in figure \ref{Histograma_Theta}, we compare the distribution of the particles positions on the sphere (in this case characterized simply by the polar angle $\theta$ and marked with the shaded bars in figure \ref{Histograma_Theta}) with the analytical solution of the diffusion equation on the sphere given by equation \eqref{Fokker-Planck-Sol} (solid-lines). 
As can be appreciated, the agreement between theoretical and numerical results is remarkable at all the times chosen. This agreement is even better than those found in \cite{Castro.2014} and in \cite{Ghosh.2012}.
We have also compared the position autocorrelation-function calculated from the position histograms as
\begin{equation}\label{AutoCorrNumerical}
   \langle \hat{\boldsymbol{n}}(t_k)\cdot \hat{\boldsymbol{n}}(0) \rangle = \frac{1}{N} \sum_{i=0}^{N} \hat{\boldsymbol{n}}_{i}(t_k)\cdot \hat{\boldsymbol{n}}_{i}(0),
\end{equation}
$N$ being the number of particles in the ensamble, with the analytical formula \eqref{AutoCorrFunct}. The agreement in this quantity is a consequence of the agreement in the position distributions (see inset in figure \ref{Histograma_Theta}).

\begin{figure}[ht]
\centering
\includegraphics[width=3.0in, trim = 0mm 0mm 0mm 5mm, clip]{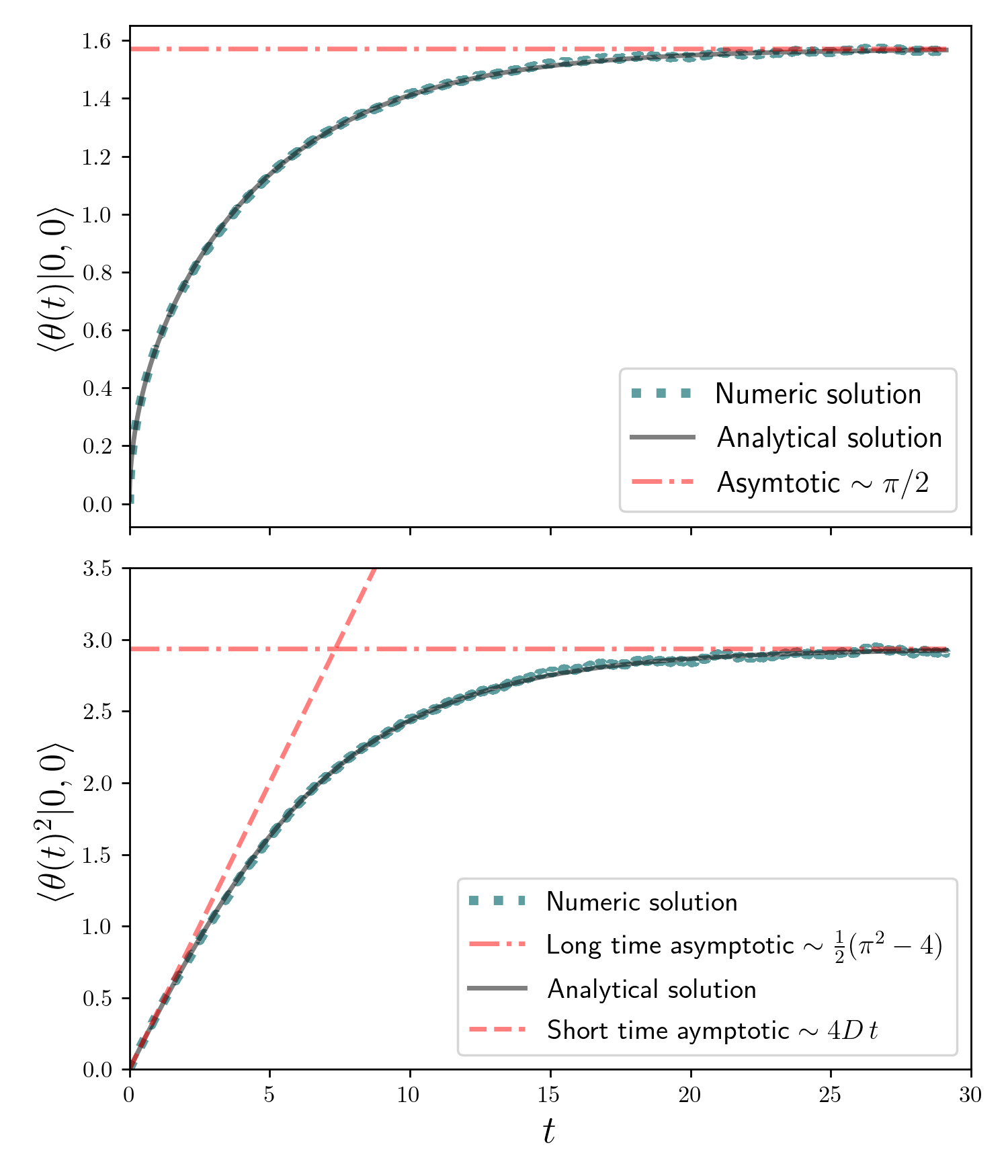}
\caption{The time dependence of $\langle\theta(t)\rangle$ (top panel), computed from the ensemble of trajectories generated with our numerical prescription (blue dots), compared with the  corresponding time dependence computed by use Eq. \eqref{Fokker-Planck-Sol} (solid-line). The dash-dotted line marks the value in the stationary regime $\pi/2$. Also, the comparison of the time dependence $\langle\theta^{2}(t)\rangle$ with both methods is shown in the bottom panel. The dashed line marks the short-time dependence, while the dash-dotted line marks the stationary regime.}
\label{Mean_Variance_Theta}
\end{figure}

Secondly, we compare the first two moments of the position distribution computed from the ensemble of trajectories obtained from our numerical algorithm, $\langle\theta(t)\rangle$, $\langle\theta^{2}(t)\rangle$, with those computed from the solution \eqref{Fokker-Planck-Sol}. The comparison is shown in figure \ref{Mean_Variance_Theta}, where a good agreement can be noticed during the transition from the initial configuration to the stationary regime.
Such agreement is reassured in the case of $\langle\theta^{2}(t)\rangle$ for which analytical formulas to compare with are known in the short- and long-time regimes. These known results have made evident where the behavior of diffusion on two dimensional \textit{curved} space ($\mathbb{S}^2$) diverges from the one of diffusion on two dimensional \textit{flat} space ($\mathbb{R}^2$). These asymptotic results can be found in \cite{Castro.2014, Castro.2014.2, Castaneda.2013} and in references therein. Explicitly, in the short-time regime, diffusion in either space must agree and (we quote the asymptotic formula)
\begin{align}\label{ShortTimeLimit}
\langle \theta^2(t) \rangle = 4 D t - \frac{2}{3} R_{g} (D t)^2 - \frac{2}{45} R_{g}^2 (D t) - \cdots,
\end{align}
where $R_g/2 = 1/r^2$ is the `Gaussian curvature'. In the long-time regime we have
\begin{align}\label{LongTimeLimit}
\langle \theta^2(t) \rangle =  \frac{\pi^2 - 4}{2} \left ( 1 - \frac{3\pi^2}{4\pi^2 - 16} e^{-2D t/r^2} + \cdots \right).
\end{align}
In figure \ref{Mean_Variance_Theta} we compare our numerical results against equations \eqref{ShortTimeLimit} and \eqref{LongTimeLimit}, as well as against the numerical integration of the analytical solution in the intermediate region. 


\subsection{Time evolution of Brownian particles in the presence of an external Field}

In this section, we analyze the transitional dynamics between two equilibrium-like stationary distributions, $\overline{P}_\text{st}^{0}(\theta,\phi)$ and $\overline{P}_\text{st}(\theta,\phi)$, provided by our numerical algorithm. Regrettably, we do not have analytical solutions of equation \eqref{SmoluOp} at the intermediate times to compare with, however this is possible to do in the stationary regime.

We start from the stationary distribution $\overline{P}_\text{st}^{0}(\theta,\phi)=\sin\theta/4\pi$ that corresponds to the stationary distribution of the positions of freely diffusing particles on the sphere. Suddenly, the external field $U_{\lambda}(\theta)$ is turned on at time $t=0$, that is, the process is described as a `quenched' dynamics induced by the time dependent external potential 
\begin{align}\label{TimePotential}
   U_{\lambda}(\theta,t) = \begin{cases} 0 & t < 0, \\ U_{\lambda}(\theta) & t \ge 0. \end{cases}
\end{align}
This sudden change perturbs the stationary dynamics and induces a response that drives the system into a new stationary regime. If the amplitude of the change is of the order of thermal fluctuations the response dynamics occurs in the \emph{linear regime} \cite{Kubo.SPII.1998}. Indeed, we are tacitly assuming that the time evolution of $\overline{P}(\theta,\phi,t)$ is driven linearly by $A^{\alpha}$. Naturally, once the field is on, the dynamics from the initial distribution $\overline{P}_\text{st}^{0}(\theta,\phi)$ occur out of equilibrium and the system will relax to a new equilibrium distribution determined by the external field. A transition between these two equilibrium distributions is then observed. In this work, our principal concern is not with this rate of approaching to equilibrium, but rather with the geometric effects of the ambient space.

After a transient relaxation, a stationary regime is reached. We show in this section that the stationary histograms obtained from our numerical algorithm agree remarkably well with stationary distributions $\overline{P}_\text{st}(\theta,\phi)$, which are solutions of the Smoluchoswski equation \eqref{SmoluOp}. Due to the azimuthal symmetry the stationary distribution depends only on the polar angle $\theta$ and is given by (see Appendix \ref{Deduc-Stationary})
\begin{equation}
\overline{P}_\text{st}(\theta,\phi)=\frac{N}{2\pi}\sin\theta\, \exp\left\{\frac{\zeta r^{2}}{k_{B}T}\int^{\theta}d\theta^{\prime}\, A^{\theta^{\prime}}_{\lambda}\right\},
\end{equation}
where we have used the Einstein relation $D\zeta=k_{B}T$. After use of equation \eqref{ForceTheta} it can be written as 
\begin{equation}\label{Stat-Sol-U}
\overline{P}_\text{st}(\theta,\phi)=\frac{N}{2\pi}\sin\theta\, \exp\left\{-\frac{U_{\lambda}(\cos\theta)}{k_{B}T}\right\},
\end{equation}
where the $N$ is the normalizing constant.  In particular, for the external fields used in this work, the corresponding stationary solutions are
\begin{subequations}\label{StatSol}
\begin{align}
\overline{P}_\text{st}(\theta,\phi) &= \frac{\lambda'\sin\theta}{4\pi k _{B}T\sinh(\frac{\lambda'}{k_{B}T})} \exp\left\{-\frac{\lambda'\, \cos\theta}{k_{B}T}\right\}, \label{ST1}  \\ 
\overline{P}_\text{st}(\theta,\phi) &= \sqrt{\frac{3\lambda'}{\pi k_{B}T}}\frac{\sin\theta}{\text{Erfi}\Bigl[\sqrt{\frac{3\lambda'}{k_{B}T}}\Bigr]} \exp\left\{\frac{3 \lambda' \cos^{2}\theta}{k_{B}T}\right\}. \label{ST2} \ 
\end{align}
\end{subequations}
where the new potential-strength factors, $\lambda'$, absorb the numerical factors that come with the spherical harmonics in the right-hand side of equation \eqref{PotentialSet}. Notice that equation \eqref{Stat-Sol-U} cannot be evaluated in terms of elementary functions when $U_{\lambda}(\theta)$ is given by equation \eqref{U3}, thus is evaluated numerically. 
\begin{figure}
\centering
\includegraphics[trim = 40mm 30mm 5mm 55mm, clip,width=3.5in]{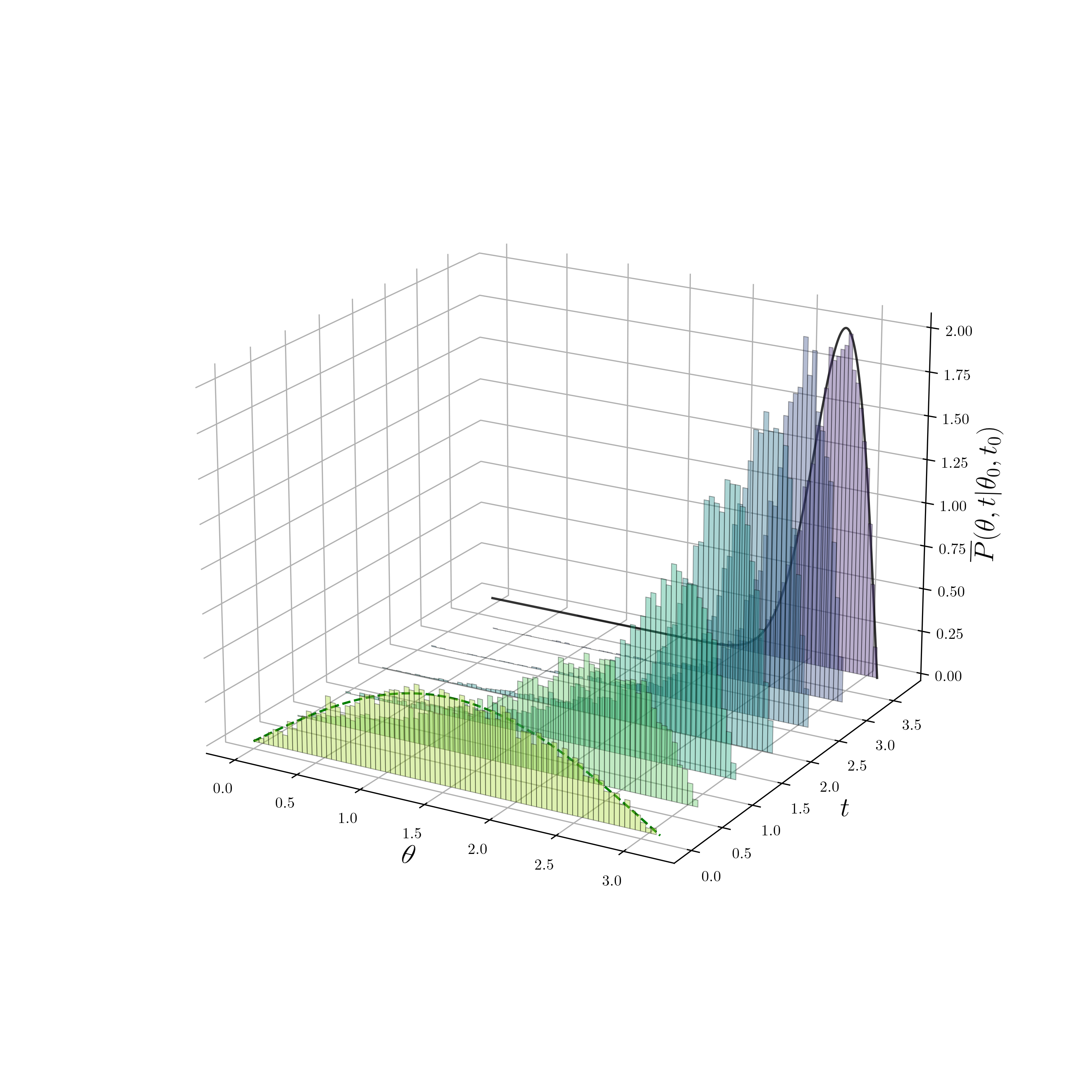}
\includegraphics[trim = 61mm 65mm 53mm 65mm, clip,width=0.35\columnwidth,height=0.35\columnwidth]{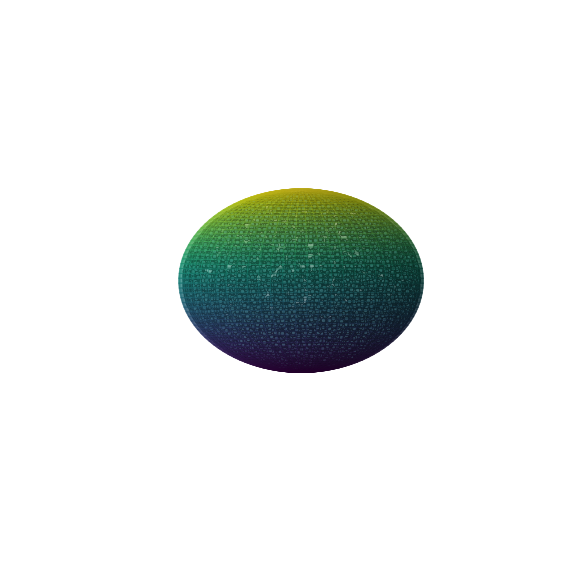}\qquad\includegraphics[trim = 61mm 65mm 53mm 65mm, clip,width=0.35\columnwidth,height=0.35\columnwidth]{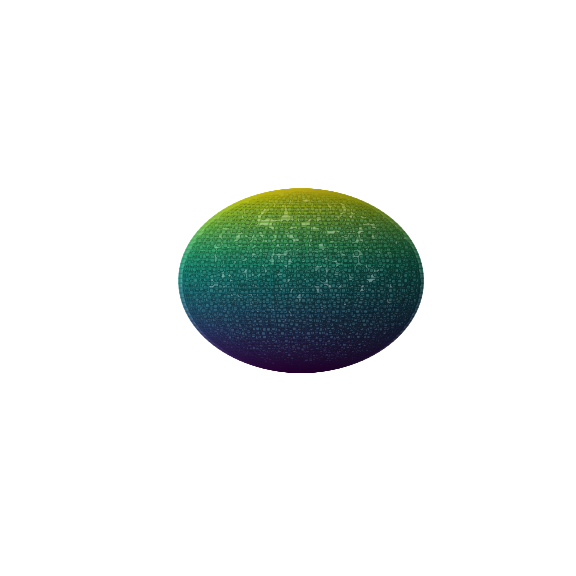}
\includegraphics[trim = 61mm 65mm 53mm 65mm, clip,width=0.35\columnwidth,height=0.35\columnwidth]{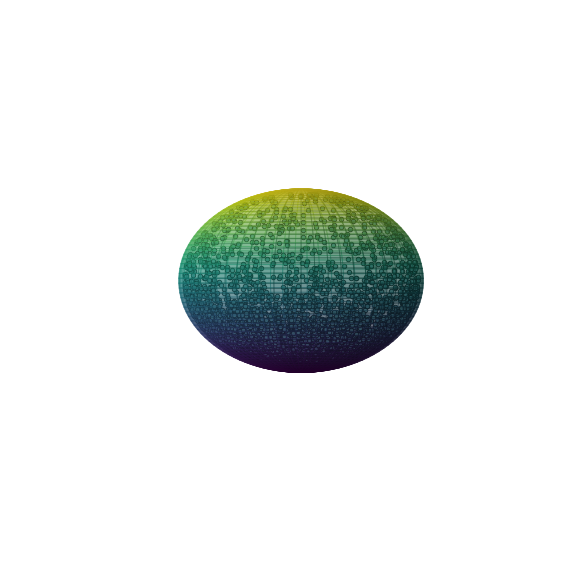}\qquad\includegraphics[trim = 61mm 65mm 53mm 65mm, clip,width=0.35\columnwidth,height=0.35\columnwidth]{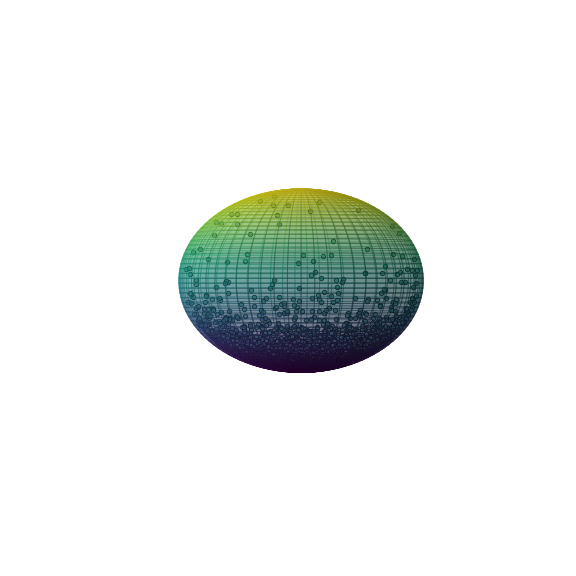}\caption{Top panel.- Histograms of $\theta$ at  times $t=0.00$, 0.62, 1.25, 1.87, 2.50, 3.12, and 3.74 obtained from an ensemble of $10^{4}$ trajectories generated with numerical prescription (shaded-color bars) for the potential \eqref{U1} $\lambda Y_{1}^{0}(\theta, \phi)$. The initial distribution (taken as explained in the main text) is indicated by a dashed green line. The corresponding stationary distribution \eqref{ST1} is marked with a solid black line. Bottom panel.- The positions of $10^{4}$ Brownian particles on the sphere are shown at times $t=0.35$, 0.69, 1.39, and 3.74, from top to bottom and from left to right. The position are obtained from our numerical algorithm and correspond to the histogram shown in the top panel. To see the animation of the complete evolution, as well as the code in Python see the complementary material of this work in GitHub \cite{GitHub_Repo}.}
\label{Sol.Num.Y10.Theta}
\end{figure}

\begin{figure}
\centering
\includegraphics[trim = 45mm 45mm 10mm 45mm, clip,width=3.5in]{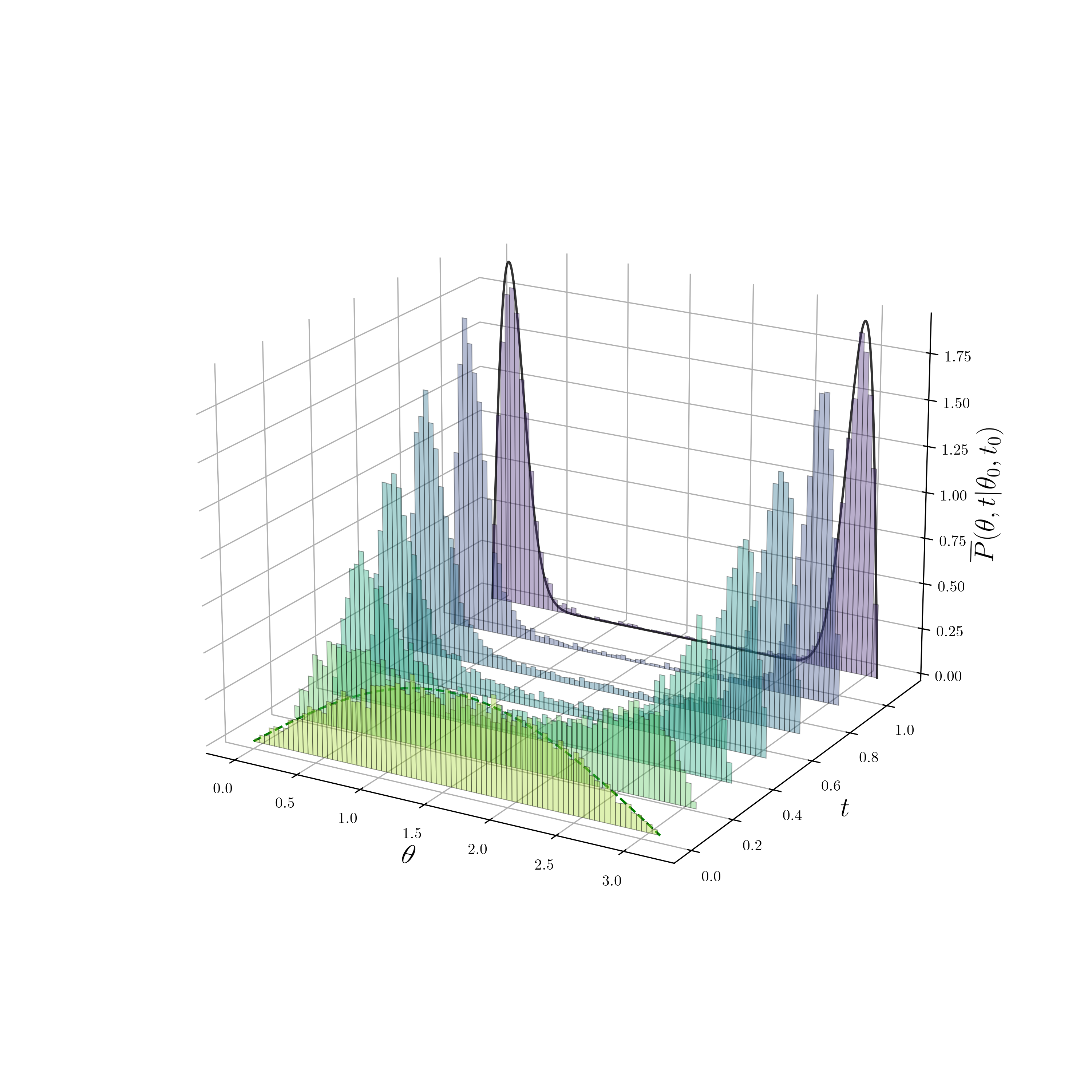}
\caption{ Histograms of $\theta$, from the initial distribution (dashed-green line), to nearly the stationary state when the interaction is given by \eqref{U2}, for five different instances of time $t=0.00$, 0.17, 0.35, 0.52, 0.69, 0.90, and 1.11. }
\label{Sol.Num.Y20.Theta}
\end{figure}

\begin{figure}
\centering
\includegraphics[trim = 45mm 45mm 10mm 45mm, clip,width=3.5in]{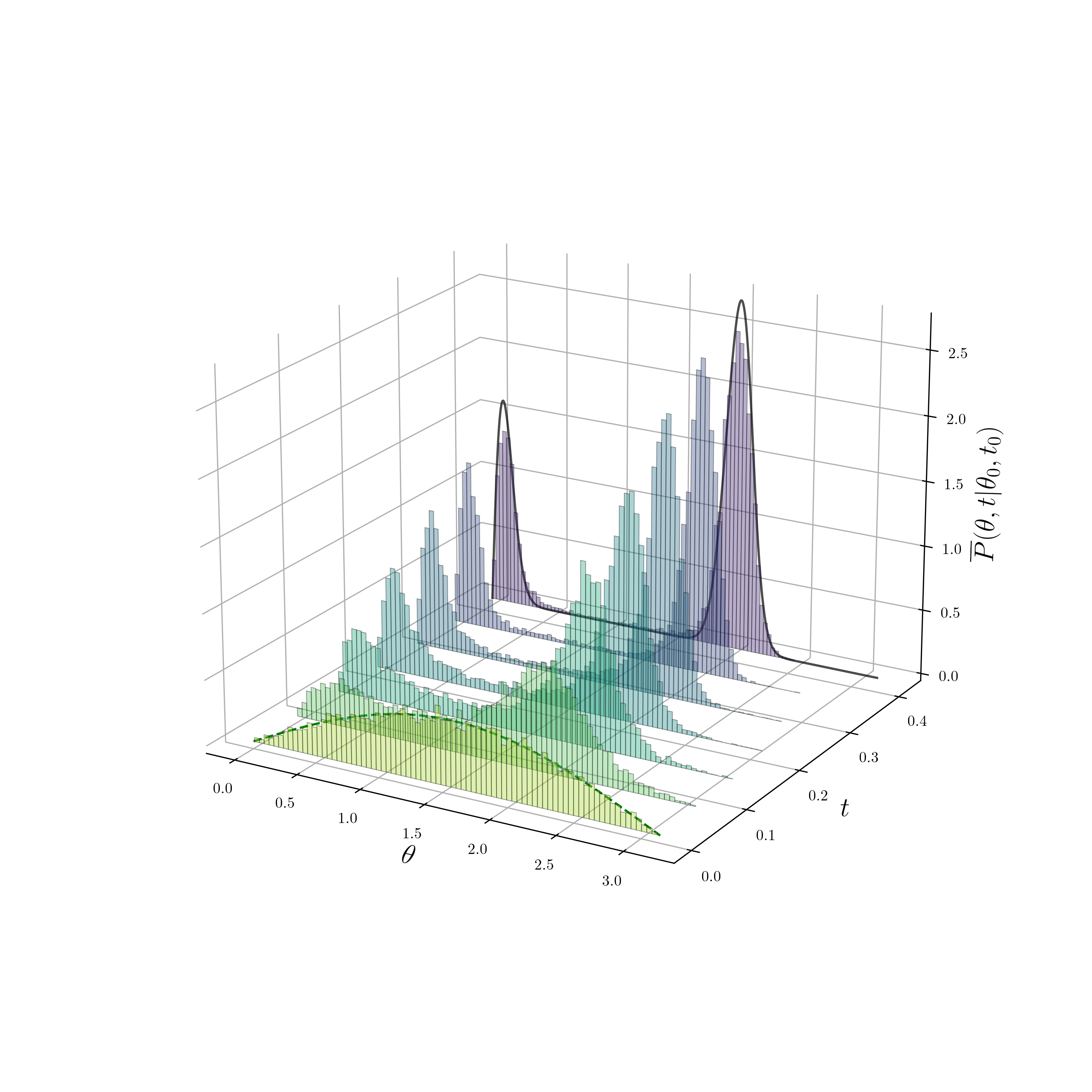}
\caption{Histogram of $\theta$, from the initial distribution (dashed-green line), to nearly the stationary state when the interaction is given by $Y_{3}^{0}(\theta, \phi)$, for five different instances of time $t=$0.00 ,0.07, 0.14, 0.21, 0.28, 0.35, and 0.42.  Analytical solution \eqref{StatSol} (histogram in purple) to the stationary Fokker-Planck equation against our numerical results, for the interaction given in \ref{U3}. This is a meta stable configuration. The true stationary state can be reached if instead of starting from a uniform initial distribution $P(\theta_0,0) = 1/2 \sin{\theta}$, we start from an infinitely concentrated distribution in the North pole $\delta(\theta_0)$.}
\label{Sol.Num.Y30.Theta}
\end{figure}

In figures \ref{Sol.Num.Y10.Theta}-\ref{Sol.Num.Y30.Theta}, the transition of the initial distribution $\overline{P}^{0}_\text{st}(\theta,\phi)$ (marked with the dashed-green line) towards the stationary distribution $\overline{P}_\text{st}(\theta,\phi)$ (solid-black line) is shown for an ensemble of $10^{4}$ trajectories computed from our numerical algorithm. figure \ref{Sol.Num.Y10.Theta} corresponds to the potential $U_{\lambda}(\theta)$ given by equation \eqref{U1}. It can be clearly noticed in the top panel that as time increases, the histograms of the particles positions (determined by $\theta$ only and marked with shaded-color bars) redistribute, peaking towards the minimum of the potential in the stationary state given explicitly by equation \eqref{ST1}. The excellent agreement of the histogram (shaded-dark bars) and the analytical solution (solid-line) is remarkable. In the bottom panel of figure \ref{Sol.Num.Y10.Theta} the distribution of the particles on the sphere is shown, from top to bottom and from left to right, at times $t=0.35$, 0.69, 1.39, and 3.74, respectively. Initially, the particles are uniformly distributed on the sphere and as time increases the potential `pushes' the particles towards the south pole.

Similarly, the time evolution of the histogram of $\theta$ (shaded-color bars) under the effects of the potential \eqref{U2} $U_{\lambda}(\theta)=\lambda Y_{2}^{0}(\theta, \phi)$ is shown in figure \ref{Sol.Num.Y20.Theta}. In this case, the external potential `pushes' the particles towards both poles until they reach a bimodal stationary distribution. Again, a good agreement between our numerical results (shaded-color bars) and  the exact stationary distribution, given by equation \eqref{ST2} and marked in the figure with solid-black line, is noticeable.

The third potential considered in our analysis \eqref{U3}, namely, $U_{\lambda}(\theta)=\lambda Y_{3}^{0}(\theta, \phi)$, has two minima: one corresponding to stable state (global minimum) and the other to a metastable one (local minimum) as was explained. The initial distribution is marked by the dashed-green-line as before, in this case, it is observed that as time passes a large fraction of the particles gets trapped in the metastable state, as is shown in figure \ref{Sol.Num.Y30.Theta} by the particle accumulation around $\theta\approx2$, while the remaining particles do distribute according the equilibrium distribution given by equation \eqref{SmoluOp}. Since the time required by the particles to escape from such state is exponentially large in the barrier height, as is established by Kramer's escape theory, the sampled distribution in our finite-time simulations differs from the corresponding equilibrium distribution solution. 

The profile of the `metastable distribution' marked by the solid-black line in figure \ref{Sol.Num.Y30.Theta} is obtained as follows: In a finite window of simulation time, the particles that initially were at angles larger than the angle that defines the barrier maximum, $\theta_\text{bmax}$, will remain in that region accumulating around the metastable state. Similarly, the particles that initially were at angles smaller than $\theta_\text{bmax}$, will remain in that region gathering around the potential global minimum.  Therefore, we introduce a `returning point' of an impenetrable potential that splits the interval $[0,\pi]$ into two unreachable regions, each one with an appropriate stationary distribution of the form \eqref{Stat-Sol-U}. The required profile is computed in such way that the fraction between the number of particles in each region stands in the same relation to the one between the areas of latitudes $[0,\theta_\text{bmax}]$ and $[\theta_\text{bmax}, \pi]$, respectively. The equilibrium distribution \eqref{Stat-Sol-U} is recovered from those initial distributions for which their time evolution is not hindered by the energy barrier. 

\begin{figure}
\centering
\includegraphics[trim = 0mm 0mm 0mm 0mm, clip,width=3.5in]{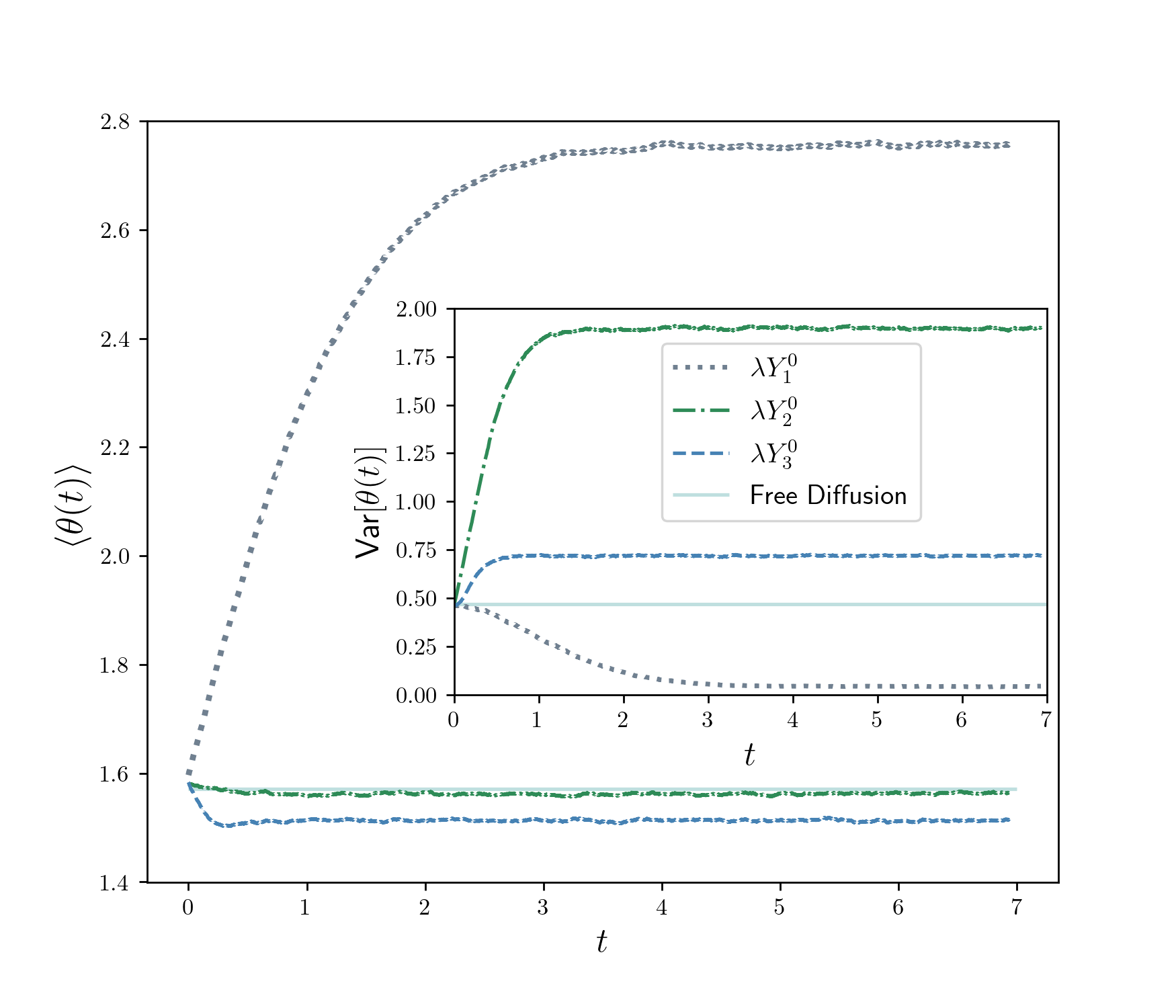}\\
\includegraphics[trim = 0mm 0mm 0mm 18mm, clip,width=3.5in]{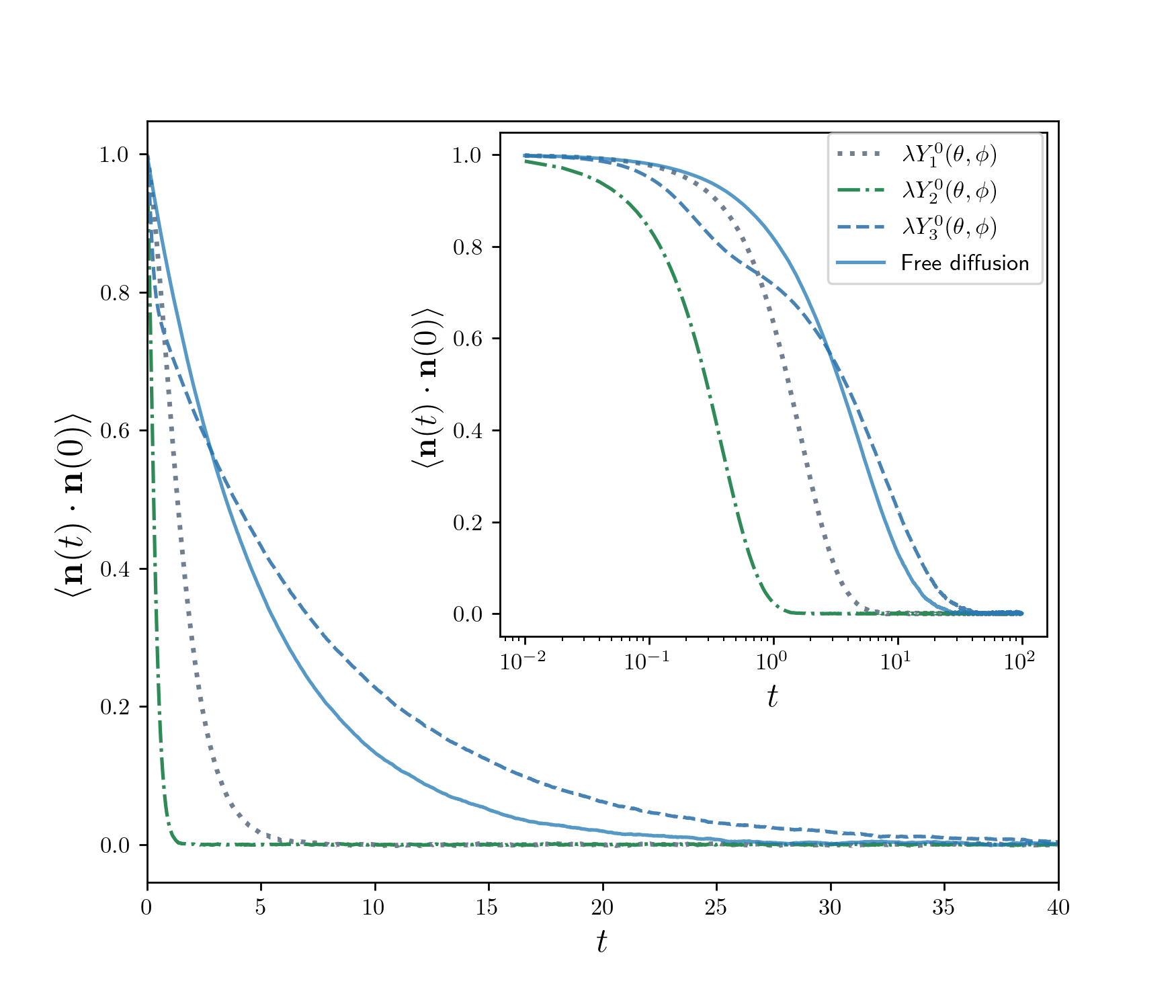}
\caption{(Top) Mean and variance (inset figure), of the  $\theta$ coordinate in the ensemble of Brownian particles, as function of time for the three different potentials in Eqs. \eqref{PotentialSet}. From this graph it can be appreciated that the \textit{relaxation time} depends on the initial distribution and the particular field. (Bottom) The autocorrelation functions for the free diffusion and the three interactions. We calculate these using an average over the ensamble of particles given by Eq. 
\eqref{AutoCorrNumerical}. The parameters used were $N=1.7\times 10^{5} $, $D=0.1$, $\lambda=1$, and $dt = 10^{-3}$.}
\label{Sol.Num.Mean_Theta}
\end{figure}

We further analyze the effects of the external potentials on the diffusion of Brownian particles on the sphere, by computing the mean position $\langle\theta(t)\rangle$, the distribution variance $\text{Var}[\theta(t)]$ (both shown in the top panel of figure \ref{Sol.Num.Mean_Theta}) and the position autocorrelation function \eqref{AutoCorrNumerical} (shown in the lower panel of the same figure). The numerical calculations for the latter case were carried out for an ensemble of $1.7\times 10^{5}$ Brownian particles.


The relaxation rate towards the stationary distribution strongly depends on the ratio $\lambda/k_{B}T$ and on the initial distribution if the external potential has metastable states. For $\lambda/k_{B}T \ll 1$, the stationary distribution $\overline{P}_{st}(\theta,\phi)$ is close to $\sin{\theta}/4\pi$ and a fast relaxation is expected. In contrast, $\overline{P}_{st}(\theta,\phi)$ is determined by the minima of $U_{\lambda}(\theta)$  if $\lambda/k_{B} T \gg 1$ and the relaxation towards the stationary distribution depends on the initial distribution. Indeed, the transition rate towards the stationary distribution is exponentially small $\exp{\{-\mbox{const}/D\}}$ if the particles has to overcome an energy barrier to reach the minimum of the potential. 


\subsection{Stability of Stationary states}

Samples of sizes $N_\text{samp}=10^6,$ $10^{7}$, $10^{8}$ and $10^{9}$, for $\theta$ in the stationary regime were generated from our numerical algorithm with integration step sizes $dt=0.001\tau$, $0.01\tau$, $0.1\tau$, in the cases of free diffusion and under the influence of the potential \eqref{U2}. Data were gathered from simulations of an ensemble of $N=10^{5}$ trajectories at times $t_n=t_{0}+n\tau$, with $n=1,\ldots,M$; $t_0$ is a large enough chosen time that it guarantees that sampling occurs in the stationary regime ($2\tau$ in the cases considered), and $\tau=r^{2}/D$ is the characteristic time scale. From these data, mean values of the quantity $f[\theta]$ are computed as
\begin{align}\label{Time-Avg}
\overline{f(t)} \equiv \frac{1}{N\cdot M} \sum_{i=1}^{N}\sum_{n=1}^{M} f[\theta_{i,n}],
\end{align}
where $\theta_{i,n}$ denotes the value of theta in the $i$-th trajectory at time $t_{n}$.

The absolute error of $\overline{f(t)}$ with respect the equilibrium value, 
\begin{align}\label{Stat-fvalue}
\langle f \rangle_\text{eq} =\int_{0}^{2\pi}d\phi \int_{0}^{\pi}d \theta\, f(\theta) \overline{P}_{st}(\theta,\phi), \end{align}
computed from the exact stationary distribution given in \eqref{Stat-Sol-U}, is given by
\begin{align}
\Delta f = \bigl\vert\overline{f(t)} - \langle f \rangle_\text{eq}\bigr\vert.
\end{align}

\begin{table}
\centering
\begin{tabular}{l c c c c c}
$dt$ & Sample & $ \Delta\theta$  &  $ \Delta \theta^{2}$   & $ \Delta \theta^{3}$ & $ \Delta \theta^{4}$  \\ \hline 
$10^{-3}$ & $10^{6} $ & 0.001696 & 0.004167  & 0.007578 & 0.011861 \\ 
& $10^{7} $ & 0.000433 & 0.000052	& 0.003786	& 0.019382	 \\ 
 & $10^{8} $ & 0.000395	& 0.001056	&0.001656 &	0.000316  \\ 
 & $10^{9} $ & 0.000014 	& $0.000008$		& 0.000042	& 0.000233	  \\ \hline
$10^{-2}$& $10^{6} $ & 0.000582 	& 0.001775 & 0.004418	& 0.010786	\\ 
& $10^{7} $	& 0.001468	&0.002569	& 0.004387 & 0.009928	 \\ 
& $10^{8} $ & 0.000020	& 0.000398	& 0.001976	& 0.007509	 \\ 
& $10^{9} $ & 0.000016	& 0.000025	& 0.000016	& 0.000066	 \\ \hline
$10^{-1}$& $10^{6} $ & 0.001791 & 0.005827	& 0.016737	& 0.046506	\\ 
& $10^{7} $ & 0.001136 & 0.005466	&0.017779	&0.052353	\\ 
& $10^{8} $ & 0.000176 & 0.002395  & 0.012649 & 0.047425	\\ 
& $10^{9}$  & $0.000005$	& 0.000020	& 0.000042	& 0.000061	\\ \hline
\end{tabular}
\caption{Absolute error, as defined in \eqref{Stat-fvalue}, of the firsts four moments of the numerically sampled distribution of $\theta$ of particles freely diffusing over the sphere surface with $D=1$. The dependence of the error on the time step size $dt$ and on the sample size is shown. }
\label{Convergence-Free}
\end{table}

\begin{table}
\centering
\begin{tabular}{l c c c c c}
$dt$ & Sample & $ \Delta\theta$  &  $ \Delta \theta^{2}$   & $ \Delta \theta^{3}$ & $ \Delta \theta^{4}$  \\ \hline 
$10^{-3}$ & $10^{6} $ & 0.001184 & 0.003358 & 0.008285  & 0.019857 \\ 
& $10^{7} $ & 0.000466	&	0.001950	&	0.0019499	&	0.065316 \\ 
 & $10^{8} $ & 0.000001	&	0.000382	&	0.001815	&	0.006381  \\ 
 & $10^{9} $ & 0.000041	&	0.000335	&	0.001822	&	0.006857  \\ \hline
$10^{-2}$& $10^{6} $ & 0.000777	&	0.006259	&	0.024629	&	0.080600 \\ 
& $10^{7} $ & 0.000166	&	0.003845	&	0.018739	&	0.065316 \\ 
& $10^{8} $ & 0.004103	&	0.004103	&	0.019810	&	0.069621 \\ 
& $10^{9} $ & 0.004282	&	0.004282	&	0.020471	&	0.071849 \\ \hline
$10^{-1}$& $10^{6} $ & 0.002058	&	0.034305	&	0.175293	&	0.628564 \\ 
& $10^{7} $ & 0.000220	&	0.040534	&	0.192456	&		0.672770 \\ 
& $10^{8} $ & 0.000005   & 0.041133 & 		0.193835 &		0.675919 \\ 
& $10^{9}$  & 0.000052	&	0.041344	&	0.194492	&	0.677735 \\ \hline
\end{tabular}
\caption{Absolute error, as defined in \eqref{Stat-fvalue}, of the firsts four moments of the numerically sampled distribution of $\theta$ of particles diffusing over the sphere surface with $D=1$ and under the influence  of the external potential \eqref{U2} with $\lambda=k_{B}T$. The dependence of the error on the time step size $dt$ and on the sample size is shown.}
\label{Convergence-Pot}
\end{table}

The absolute error for the firsts four moments of $\overline{P}_\text{st}(\theta,\phi)$, namely $\Delta\theta$, $\Delta\theta^{2}$, $\Delta\theta^{3}$ and $\Delta\theta^{4}$ is presented in table \ref{Convergence-Free} for the case of free diffusion, and in table \ref{Convergence-Pot} of diffusion under the influence of the external potential \eqref{U2}. 

As expected, for smaller integration step sizes, the error is rather sensible to sample sizes as is observed in the firsts four moments considered for $dt=0.001$. It can also be noticed that for a chosen integration step size, the error diminishes as the sample size gets large. This trend, however, seems to stops at sample sizes between $10^{8}-10^{9}$, thus sample size does not play an important role in error for large enough samples. This effect is observed even earlier for high moments, for instance, no improvement of the error is found in the fourth moment by enlarging the sample size when $dt=0.01$. From these results, we conclude that this analysis gives a measure the intrinsic error of our numerical algorithm.       

Finally, the tolerated error that we may allow ourselves from a given numerical method depends on the particular problem under consideration. Indeed, more exact results might be generated at the expense of more demanding computational power. The analysis presented here might serve to have a starting point to see whether the numerical algorithm may be feasible for a particular problem with a specific associated tolerance.


\section{Summary and concluding remarks}
\label{Discuss}

In this paper, we presented a numerical algorithm to generate the trajectories of Brownian particles diffusing on the surface of the two-dimensional sphere. The algorithm implemented is based on three-dimensional geometry, thus avoiding the complications introduced by the interpretation of the multiplicative Gaussian-white noise involved in the corresponding stochastic differential equation, and takes into account the effects of an external potential. Our numerical method is weakley convergent, that is, it statistically converges to the solution of the Smoluchowski equation (\ref{SmoluOp}).

The algorithm was validated in two scenarios: First, in the absence of external potential, by computing the time dependence histogram of the polar angle, as well as the first two moments, and the time correlation function of the particle
positions, from an ensemble of trajectories generated by our algorithm. We observed an excellent agreement at all times between our results and the corresponding quantities obtained from the solution to the diffusion equation on the sphere (\ref{Fokker-Planck-Sol}). Second, in the presence of a external potential, for which we calculated the stationary probability distribution when the external potential depends only on $\cos{\theta}$. Again, excellent agreement was found between our numerical results and those given by the stationary solutions of the Smoluchoswki equation (\ref{SmoluOp}) for three different potentials. In addition, we have also calculated the time correlation of the particle positions for each of these three potentials. Although no analytical solution are available to our knowledge, the traits of correctness are observed. Although we have considered simple potentials, our algorithm can be applied to an arbitrary but physically motivated potential.

An alternative to our numerical algorithm might consider either the numerical integration of the differential stochastic equations that determine the time evolution of $\theta(t), \varphi(t) $ or the numerical solution of the Smoluchowski equation (\ref{SmoluOp}). However, in any case, this procedure requires to deal with singular terms that diverge at the poles. Although this can be solved by changing from one set of coordinates (chart) to another one, this is however, computationally time consuming.


\acknowledgments
We acknowledge to the anonymous referees for their thorough   and carefully review of the original manuscript. This work was supported by UNAM-PAPIIT IN110120. A.V.G kindly acknowledges the scholarship received from CONACyT and  UNAM-PAPIIT IN110120.

\appendix


\section{Transformation of the Fokker-Planck equation}
\label{Trans-Fokk}

The Fokker-Planck equation in Cartesian coordinates  is
\begin{align}\label{Fokker-Euc}
   [P],_{t}= -  [A^{k} P],_{k} + \frac{1}{2} [B^{k l} P],_{k l}
\end{align}
If we make a coordinate transformation $x^{i} \to \eta^{\alpha}$ the chain rule implies that this equation transforms accordingly
\begin{align}\label{Transformed_Fokker-Planck}
[ P],_{t}  &=  - [ A^{k} P ],_{\alpha} \Lambda^{\alpha}_{k} \nonumber \\
&+ \frac{1}{2} \left \{   [B^{k l} P],_{\alpha \beta} \Lambda^{\alpha}_{k} \Lambda^{\beta}_{l} + [B^{k l} P],_{\alpha} \Lambda^{\alpha}_{k},_{l}     \right \}. \
\end{align}
However, the conditional probability density transforms as $\overline{P} = J P $ where $J$ is the determinant of the Jacobian matrix. We would like to rewrite this equation in terms of $\overline{P}$. For this purpose we need some elementary results from linear algebra. We know that the Jacobian matrix $\Lambda^{\alpha}_{k} = \partial \eta^{\alpha} / \partial x^{k}$ and its inverse $\Lambda^{k}_{\beta} = \partial x^{k}/\partial \eta^{\beta}$, satisfy
\begin{align}
\Lambda^{\alpha}_{k} \Lambda^{k}_{\beta} = \delta^{\alpha}_{\beta}
\end{align}
so if $C^{j}_{\alpha}$ is the cofactor associated with the term  $\Lambda^{\alpha}_{j}$, then this is given by
\begin{align}
C^{\alpha}_{j} = J' \Lambda^{\alpha}_{j},
\end{align}
where $J'$ is the Jacobian determinant of the inverse matrix $\Lambda^{k}_{\beta} $. Using this relation we can express the determinant derivative with respect the elements $\Lambda^{\alpha}_{j}$, in the following manner
\begin{align}
J',_{\Lambda^{\alpha}_{j} } = C^{j}_{\alpha} = J' \Lambda^{j}_{\alpha}.
\end{align}
Now we observe that the partial derivatives of the Jacobian determinant can be related to the components of the Jacobian matrix
\begin{align*}
-J^{-1} J,_{i} = -[ \ln{J}],_{i} = [\ln{J'}],_{i} = J'^{-1} J',_{i}  =  & \\  
J'^{-1} J',_{\Lambda^{\alpha}_{j}} \Lambda^{\alpha}_{j},_{i} = \Lambda^{j}_{\alpha} [ \Lambda^{\alpha}_{j} ],_{i} 
= \Lambda^{j}_{\alpha}[ \Lambda^{\alpha}_{i}    ],_{j} = \Lambda^{\alpha}_{i} ,_{\alpha}. &  \
\end{align*}
Then we can express the derivatives with respect the old coordinates $x^{i}$, as functions of the new coordinates  $x^{\alpha}$ in the following manner
\begin{align}\label{newcoord_Jacob}
[\cdot],_{i} = \Lambda^{\alpha}_{i} [\cdot],_{\alpha} = [\Lambda^{\alpha}_{i} \cdot ],_{\alpha} - \Lambda^{\alpha}_{i},_{\alpha} \cdot = [\Lambda^{\alpha}_{i} \cdot ],_{\alpha}  + J^{-1} J,_{i} \cdot \, .
\end{align}
Furthermore, it always holds
\begin{align}
[\cdot],_{i} = J^{-1} [J \cdot],_{i} - J^{-1} J,_{i}\cdot ,
\end{align}
using the relation \ref{newcoord_Jacob} we obtain a central result 
\begin{align}\label{firstDeriv-new}
[\cdot ],_{i} = J^{-1}  [ \Lambda^{\alpha}_{i} J  ],_{\alpha}.
\end{align}
Compounded twice
\begin{align*}
[\cdot],_{ij} &= J^{-1}  [ \Lambda^{\alpha}_{i} J  [  J^{-1}  [ \Lambda^{\beta}_{j} J  ],_{\beta} ]   ],_{\alpha} = J^{-1}  [ \Lambda^{\alpha}_{i}    [ \Lambda^{\beta}_{j} J  ],_{\beta}   ],_{\alpha} \\
&= J^{-1} \left \{  [\Lambda^{\alpha}_{i}],_{\alpha} [\Lambda^{\beta}_{j} J],_{\beta} + \Lambda^{\alpha}_{i}[\Lambda^{\beta}_{j} J],_{\beta \alpha}   \right \}.
\end{align*}
But at the same time
\begin{align*}
J^{-1} [\Lambda^{\alpha}_{i} \Lambda^{\beta}_{j} J \cdot ],_{\alpha \beta} 
- J^{-1} [[\Lambda^{\alpha}_{i} \cdot ],_{\beta} \Lambda^{\beta}_{j} J \cdot ],_{\alpha}  & \\ 
=  J^{-1} [\Lambda^{\alpha}_{i} \Lambda^{\beta}_{j} J \cdot ],_{\alpha \beta}
 - J^{-1} [\Lambda^{\alpha}_{i},_{j} J \cdot],_{\alpha} = J^{-1} \left \{  [\Lambda^{\alpha}_{i} ],_{\alpha \beta} \Lambda^{\beta}_{j} J \cdot \right.  & \\
 \left. +  [\Lambda^{\alpha}_{i}],_{\beta}[ \Lambda^{\beta}_{j} J \cdot],_{\alpha}  +  [\Lambda^{\alpha}_{i} \cdot],_{\alpha} [\Lambda^{\beta}_{j} J \cdot],_{\beta}  + \Lambda^{\alpha}_{i} [ \Lambda^{\beta}_{j} J \cdot],_{\alpha \beta} \right \} & \\
- J^{-1} \left \{ [\Lambda^{\alpha}_{i}],_{\alpha \beta} \Lambda^{\beta}_{j} J \cdot  +  [\Lambda^{\alpha}_{i} \cdot],_{\beta} [\Lambda^{\beta}_{j} J \cdot],_{\alpha}   \right \} & \\
= J^{-1} \left \{ [\Lambda^{\alpha}_{i} \cdot],_{\alpha} [\Lambda^{\beta}_{j} J \cdot],_{\beta} + \Lambda^{\alpha}_{i} [\Lambda^{\beta}_{j} J \cdot],_{\alpha \beta} \right \} .  \
\end{align*}
Therefore
\begin{align}\label{secondDeriv-new}
[\cdot],_{ij}  = J^{-1} [\Lambda^{\alpha}_{i} \Lambda^{\beta}_{j} J \cdot ],_{\alpha \beta} - J^{-1} [ \Lambda^{\alpha}_{i},_{j} J \cdot ],_{\alpha}.
\end{align}
Using  \ref{firstDeriv-new} and \ref{secondDeriv-new}  in the Fokker-Planck equation \ref{Fokker-Euc}
\begin{align}
   [P],_{t}= -   J^{-1}  [ \Lambda^{\alpha}_{i}  A^{i} J P ],_{\alpha} + \frac{1}{2} J^{-1} [\Lambda^{\alpha}_{i} \Lambda^{\beta}_{j}  B^{i j} J P ],_{\alpha \beta}\nonumber \\
    - \frac{1}{2} J^{-1} [ \Lambda^{\alpha}_{i},_{j}  B^{i j} J P ],_{\alpha}.
\end{align}
What allow us to write, if we gather the first derivatives in the same term, the Fokker-Planck in the new coordinates as
\begin{align}
[\overline{P}],_{t}= -  [A^{\alpha} \overline{P}],_{\alpha} + \frac{1}{2} [B^{\alpha \beta} \overline{P}],_{\alpha \beta}.
\end{align}
This implies the following transformation law for the elements of the Fokker-Planck, in terms of the elements described in the old coordinates $x^{i}$.
\begin{align}
\overline{P} &= P J = P(\eta^{\alpha}, t | \eta^{\alpha}_0, t_0) \det{ \Lambda^{i}_{\alpha} }  \\
A^{\alpha} &= \Lambda^{\alpha}_{i} A^{i} + \frac{1}{2} \Lambda^{\alpha}_{i},_{j} B^{i j} \\
B^{\alpha \beta} &= \Lambda^{\alpha}_{i} \Lambda^{\beta}_{j}  B^{i j}. \
\end{align}
Just the diffusion matrix $B^{i j}$, transforms as a (second rank) tensor; the drift vector $A^{k}$ and the conditional probability density $P$, do not transform as tensors.



\section{Stationary sates on $ \mathbb{S}^2$ }\label{Deduc-Stationary}

In this section we deduce the stationary solutions of the Fokker-Planck equation. The procedure is discussed in \cite{Haken-Graham-1971} or in \cite{Stratotonovich.2014}. The existence of the stationary solution $\overline{P}_\text{st}$ solution can guaranteed when the conditions of \textit{detailed balance} are met. In those cases we can get an explicit solution 
\begin{align}\label{CDB-Stat}
0 &= -  (D^{\alpha} \overline{P}_\text{st}),_{\alpha} + \frac{1}{2} (B^{\alpha \beta} \overline{P}_\text{st}),_{\alpha \beta} \nonumber \\
&= -  [  D^{\alpha} \overline{P}_\text{st} -  \frac{1}{2}  (B^{\alpha \beta} \overline{P}_\text{st}),_{\beta} ],_{\alpha}  \
\end{align}
in which $\overline{P}_\text{st}(\eta)=\int P(\mathbf{\eta} ,t | \mathbf{\eta}_0 ,t_0) \, d \mathbf{\eta}_0 $. The drift $D^{\alpha}$ in \ref{CDB-Stat}, as is explained in \cite{Haken-Graham-1971} is defined as $D^{\alpha} := 2^{-1}[A^{\alpha} + \tilde{A}^{\alpha}]$, which is the \textit{reversible} part of the drift \cite{Graham.1971}, but in the cases with which we dealt with, it coincides with $A^{\alpha}$ itself because it is invariant with respect to \textit{time inversion} (velocities and magnetic fields are not). If we take
\begin{align}
\overline{P}_\text{st}( \mathbf{\eta}  )= N \exp{[-U^{e}_{\lambda}( \mathbf{\eta} )]} \label{StationarySol}
\end{align}
$N$ being a normalization constant, and $U^{e}_{\lambda}$ plays the role of a \textit{generalized thermodynamic potential}; these functions are defined even far from thermal equilibrium like in a laser. If we substitute Eq. \eqref{StationarySol} in Eq. \eqref{CDB-Stat}, we obtain
\begin{align*}
D^{\alpha} N \exp{ [- U^{e}_{\lambda}(\mathbf{\eta})] } - \frac{1}{2}  B^{\alpha \beta},_{\beta} N \exp{ [- U^{e}_{\lambda}( \mathbf{\eta}  )]} =& \\
 -\frac{1}{2}  B^{\alpha \beta}  U^{e}_{\lambda},_{\beta} N \exp{[- U^{e}_{\lambda} (\mathbf{\eta}  )]}. & \
\end{align*}
If the matrix $B^{\alpha \beta }$ has an inverse, that we denote by $(B^{-1})_{\alpha \beta}$,  then we can solve for $U^{e}_{\lambda},_{\beta}$
\begin{align}
U^{e}_{\lambda},_{\alpha} = (B^{-1})_{\alpha \beta} \left [  B^{\beta \gamma},_{\gamma} - 2 D^{\beta}   \right  ] =: G_{\alpha}. \ 
\end{align}
Moreover, we may solve for $U^{e}_{\lambda}$, and reduce the problem to quadratures, if the \textit{potential conditions} \cite{Stratotonovich.2014} are satisfied
\begin{align}
G_{\alpha},_{\delta} =  \left [(B^{-1})_{\alpha \beta}   B^{\beta \gamma},_{\gamma} - 2 (B^{-1})_{\alpha \beta} D^{\beta}   \right  ],_{\delta} &= \nonumber \\
\left [(B^{-1})_{\delta \beta}  B^{\beta \gamma},_{\gamma} - 2 (B^{-1})_{\delta \beta} D^{\beta}   \right  ],_{\alpha} = G_{\delta},_{\alpha}. & \
\end{align}
That is, we might be able to obtain $U$ as a line integral over the configuration space variables $\{ \mathbf{\eta} \}$. So the particular solution for the cases treated in this work is \ref{StationarySol}, where $U^{e}_{\lambda}$ is given by
\begin{align}\label{GPot-Int}
U^{e}_{\lambda}(\theta) &=  -  \int^{\theta}  2 (B^{-1})_{\theta \theta} \left[  \Lambda^{\theta}_{k}A^{k}_{\lambda}[\eta]- \frac{D}{r^2} \cot{\eta} \right] \,d \eta  , \
\end{align}
where $U^{e}_{\lambda}$ stands for \textit{effective potential}.


\end{document}